\documentclass[final,3p,onecolumn]{elsarticle}
\biboptions{super}
\setcitestyle{open=,close=} 
\biboptions{sort&compress} 
\usepackage[table]{xcolor}
 \usepackage{multirow}
\definecolor{myred}{rgb}{0.7961, 0.2627, 0.2039}  %{203, 67, 52} 
\definecolor{mygreen}{rgb}{0.1294, 0.6039, 0.3294}  %{33, 154, 84} 
\usepackage{booktabs}
\usepackage{amssymb}
\usepackage{amsmath}
\usepackage[colorlinks=true, citecolor=blue]{hyperref}   % makes refs and citations clickable
\usepackage{cleveref}   % makes refs smart (adds "Figure", "Table", etc.)
\usepackage{booktabs} 
\usepackage{textcase}

\usepackage{xcolor}
\journal{JSV}
\begin{document}
\begin{frontmatter}
\title{\large{\textbf{Metamaterial-Inspired Bi-resonators Vibration Absorbers for Railway Tracks: Experimental Study of Flexural Wave Control}}}

\author[label1]{Ali Jafari}
\author[label2]{Aref Afsharfard\corref{mycorrespondingauthor}}\cortext[mycorrespondingauthor]{Corresponding author}
\ead{afsharfard@pusan.ac.kr}

\affiliation[label1]{organization={Dept. of Mechanical and Aerospace Engineering, University at Buffalo (SUNY)},
            city={Buffalo},
            postcode={14260-4400},
            state={NY},
            country={USA}}

\affiliation[label2]{organization={Eco-Friendly Smart Ship Parts Technology Innovation Center, Pusan National University},
            city={Busan},
            postcode={46241},
            state={NY},
            country={Republic of Korea}}

% --- Author Block ---
% \author{Ali Jafari$^{1}$}
% \author{Aref Afsharfard$^{2}$}
% \altaffiliation[Corresponding author: ]{afsharfard@pusan.ac.kr}

% \affiliation{\vspace{2ex} \makebox[\linewidth][c]{{\normalfont $^1$}\footnotesize{Dept. of Mechanical and Aerospace Engineering, University at Buffalo (SUNY), Buffalo, NY 14260-4400, USA}}\\
% \makebox[\linewidth][c]{{\normalfont $^2$}\footnotesize{Eco-Friendly Smart Ship Parts Technology Innovation Center, Pusan National University, Busan 46241, Republic of Korea}}\\
% }

\begin{abstract}
Subway rail vibrations are a major source of structural deterioration, environmental noise, and passenger discomfort in urban railway systems. Here, we present a broadband design methodology for railway tuned mass dampers (TMDs) based on the concept of joining multiple locally resonant bandgaps. The proposed framework begins with an experimental modal analysis of a UIC60/60E1 rail to identify its dominant vibration modes and develop an equivalent dynamic model. Guided by the proposed bandgap-joining strategy, single- and multi-resonator TMDs are subsequently designed, fabricated, and experimentally validated before being implemented on the railway track. The experimental investigation demonstrates that the tuned single-resonator configuration reduces the vibration amplitudes at the dominant resonances by up to 74\%, while incorporating a second resonator further broadens the effective attenuation bandwidth, confirming the advantages of the proposed multi-resonator concept. Finally, a random vibration analysis is performed to evaluate the effectiveness of the designed TMDs under stochastic excitations representative of practical railway operating conditions, predicting an average reduction of approximately 12\% in the RMS vibration response. The proposed methodology provides a practical framework for translating locally resonant metamaterial concepts into compact, manufacturable, and non-invasive railway vibration absorbers with enhanced broadband vibration mitigation capabilities.
\end{abstract}


\end{frontmatter}

%%%%%%%%%%%%%%%%%%%%%%%%%%%%%%%%%%%%%%
\section{Introduction 
\label{sec:Intro}}
Railway transportation has become an indispensable component of modern urban infrastructure because of its high carrying capacity, energy efficiency, and environmental sustainability. As urban populations continue to increase, railway systems are required to operate at higher speeds and frequencies while maintaining strict standards for passenger comfort, infrastructure reliability, and environmental protection. Despite these advantages, the dynamic interaction between the wheel and rail generates significant vibrations that propagate through the track structure and into the surrounding environment. These vibrations contribute to ground-borne noise, accelerate the deterioration of railway components, reduce passenger comfort, and may adversely affect nearby buildings and sensitive facilities such as hospitals, laboratories, and residential areas\cite{connolly2016growth, connolly2015benchmarking, kurzweil1979ground}. Consequently, effective vibration mitigation has become a critical aspect of railway system design and maintenance\cite{ouakka2022railway}.

The vibration characteristics of railway tracks are inherently complex because the rail behaves as a continuous elastic structure with multiple flexural resonances distributed over a broad frequency range \cite{xia2010theoretical}. The frequencies and amplitudes of these resonances are influenced by several factors, including the rail profile, support stiffness, sleeper spacing, fastening systems, and operational loading conditions. As a result, vibration energy is rarely concentrated at a single frequency but is instead distributed across multiple resonant modes. This broadband dynamic behavior presents a significant challenge for vibration mitigation strategies, since suppressing one resonance often has little influence on neighboring modes or may even shift vibration energy to other frequency regions.

Designing an effective vibration mitigation system first requires a thorough understanding of the dynamic behavior of the railway track. Because the rail behaves as a continuous structure with multiple interacting vibration modes, accurately identifying its dominant resonances is essential for determining the frequencies that should be targeted by vibration control devices. To this end, both operational modal analysis (OMA) and experimental modal analysis (EMA) have been widely employed to characterize the dynamic properties of railway structures\cite{Yang2021,Anastasopoulos2021}. OMA estimates modal parameters directly from operational responses under ambient loading conditions, whereas EMA utilizes controlled excitations to obtain frequency response functions (FRFs), from which natural frequencies, damping ratios, and mode shapes can be extracted \cite{Hou2025,Afsharfard2023,Panda2024}. These experimentally identified modal properties have been extensively used to construct reduced-order dynamic models of railway structures and to validate numerical simulations of rail vibration and wave propagation \cite{Aizpun2014,Lou2005,Zhang2021,Thompson1997,Oregui2015}.

Once the dynamic characteristics of the railway have been identified, the next challenge is mitigating the resulting vibrations. Numerous passive and active vibration control techniques have been proposed for railway systems, including under-sleeper pads, resilient fasteners\cite{Jin2020a,Ouakka2022}, ballast reinforcement\cite{Chi2024}, suspension optimization and vibration absorbers\cite{Jin2020b}. While modifications to the railway infrastructure can effectively reduce vibration, they often require substantial construction, increased maintenance, or high implementation costs. Consequently, passive vibration absorbers have attracted considerable attention because they can be retrofitted to existing railway systems with minimal structural modifications while providing effective attenuation around their target frequencies \cite{Karimmirza2024,SepehriAmin2020,Wang2015,Mohanty2021}.

Among the various passive vibration control technologies, tuned mass dampers (TMDs) have emerged as one of the most practical and effective solutions because of their simple mechanical design, ease of retrofitting, and ability to attenuate structural resonances without requiring modifications to the existing railway infrastructure\cite{Zhang2021b,jafari2022conceptual,mousavi2023disturbance}. TMDs have been successfully applied to railway bridges, rail vehicles, and track structures for mitigating vibrations induced by train passages, seismic events, and other dynamic loads \cite{Chen2019,Andersson2015,Liu2018,Chen2019b,mazloom2025innovative}. Their effectiveness has been further demonstrated in high-speed railway applications, where multiple tuned mass dampers (MTMDs) have been employed to reduce structural displacements, vibration amplitudes, and noise radiation \cite{Li2005,Zhang2024,Sun2019,Wu2008}. Despite these advantages, conventional TMDs remain inherently narrowband devices, providing significant vibration attenuation only within a limited frequency range centered around their tuning frequency. Since railway tracks exhibit multiple closely spaced flexural resonances over a broad frequency spectrum, a single TMD cannot effectively suppress all dominant vibration modes. Although MTMD systems extend the attenuation bandwidth by employing multiple independently tuned absorbers, they generally require a larger installation footprint, increased structural complexity, and additional mass, limiting their practicality for compact railway vibration mitigation systems.

Recent advances in locally resonant acoustic and elastic metamaterials have demonstrated an alternative approach for achieving broadband wave attenuation\cite{ma2016acoustic, zhou2012elastic}. Unlike conventional vibration absorbers that target individual resonances, locally resonant metamaterials generate frequency bandgaps through the interaction between the host structure and embedded resonant elements, enabling efficient suppression of flexural waves over prescribed frequency ranges \cite{gorshkov2021multi,aladwani2020mechanics,iqbal2021flexural}. Furthermore, broadband attenuation can be achieved by combining or overlapping multiple locally resonant bandgaps, thereby extending the effective frequency range without substantially increasing the overall dimensions of the structure\cite{meng2023theoretical}. Despite these advances, the underlying principles of locally resonant bandgap engineering have rarely been translated into practical railway vibration absorbers. In particular, a systematic methodology for incorporating multiple resonant elements within a compact tuned mass damper to realize broadband vibration suppression has not yet been established.

In addition to the vibration behavior of tracks, the concept of power spectral density (PSD) is widely applied in railway dynamics to evaluate the influence of random track irregularities and induced vibrations on moving vehicles \cite{Radgolchin2018, Reumers2025}. Vibration-absorbing systems also have a close correlation with the random vibration concept, as they are efficient tools in suppressing stochastic displacements \cite{Barredo2023}, and manufacturing errors and operating environments of TMD lead to uncertainties in acoustic and vibration performance \cite{Zhang2025}. As shown by Xu et al. \cite{Xu2021}. PSD is combined with train loads to create dynamic models for stochastic analysis in train-track interactions. To enhance bridge behavior predictions under operational conditions, Gadysz and Niad \cite{Gladysz2009} carried out a spectral density analysis of beam bridge vibrations, considering stochastic variability in train-induced loads and structural response parameters. Using stochastic modeling to evaluate the response of high-speed railway bridges to random ground excitations, Zhu et al. \cite{Zhu2022} carried out a thorough investigation into the seismic performance of these structures. Liu et al. \cite{Liu2020} developed a random analysis procedure for railway vehicles using pseudo-excitation theory to assess random vibrations in railway vehicles. To better understand the dynamic interactions between train-bridge coupled systems under non-uniform earthquake conditions, Zhao et al. \cite{Zhao2023b} proposed a stochastic methodology. To assess and lessen the effects of sporadic vibrations and seismic activity on railway infrastructure and help create more robust and dependable railway systems, these collaborative efforts highlight the significance of stochastic approaches.

Motivated by this research gap, this work proposes a broadband design methodology for railway tuned mass dampers based on the concept of joining multiple locally resonant bandgaps within a single absorber. The proposed framework first characterizes the dynamic behavior of the railway track through experimental modal analysis and reduced-order modeling, followed by the systematic design of single- and multi-resonator TMDs according to the proposed bandgap-joining strategy. The designed absorbers are experimentally validated on a railway track using a finite periodic arrangement, and their performance is further evaluated under stochastic excitation through random vibration analysis. Unlike conventional MTMD approaches that distribute multiple independent absorbers along the structure, the proposed methodology integrates multiple locally resonant elements within each absorber, providing a compact and practical solution for broadband railway vibration mitigation. The remainder of this paper is organized as follows. Section~2 presents the theoretical framework for achieving broadband vibration attenuation through the joining of locally resonant bandgaps and establishes the design principles of the proposed multi-resonator concept. Section~3 translates this framework into the design of practical railway TMDs, beginning with the experimental characterization and modal modeling of the railway track, followed by the design methodology of the multi-resonator TMDs and the experimental validation of the fabricated resonators. Section~4 experimentally evaluates the proposed TMD configurations on a railway track, investigating the performance of both single-resonator and bi-resonator designs using a finite periodic arrangement of absorbers. Section~5 presents a random vibration analysis to assess the effectiveness of the proposed TMDs under stochastic excitation representative of realistic railway operating conditions. Finally, the main conclusions of the study are summarized in Section~6.
%%%%%%%%%%%%%%%%%%%%%%%%%%%%%%%%%%%%%%%%%%%%%%%
\section{Evidence of joining locally resonant bandgaps}
\label{joining_bandgaps}
Locally resonant acoustic and elastic metamaterials are well known for their ability to generate low-frequency bandgaps through subwavelength resonators embedded in, or attached to, a host medium. In classical single-resonator configurations, the resulting locally resonant bandgap is typically narrow and centered around the natural frequency of the resonator. While this mechanism is effective for targeted vibration attenuation, it limits the achievable bandwidth and restricts the flexibility of the design when broadband suppression is desired. Prior to advancing into the modeling framework required for developing the design tools that form the core of this article, this section will elucidate an intriguing emergent phenomenon in dissipative metamaterials, commonly known as joining locally resonant bandgaps \cite{lewinska2017attenuation, meng2023theoretical, sugino2018merging, guo2025bandgap}. 

A natural strategy to overcome this limitation is to introduce multiple local resonators with distinct natural frequencies within the same unit cell. In principle, each resonator contributes its own locally resonant bandgap, suggesting the possibility of broadband attenuation through the superposition of multiple gaps. However, in practice, the interaction between resonators and the host structure leads to nontrivial coupling effects, and the resulting bandgaps do not simply combine in an additive manner. As a result, the conditions under which adjacent locally resonant bandgaps merge into a single, continuous stop band remain insufficiently understood. The key idea is to tune the mass and frequency ratios of two or more internal resonators such that the upper bound of the first locally resonant bandgap coincides with, or overlaps, the lower bound of the second bandgap. When this condition is satisfied, the intermediate pass band vanishes, yielding a continuous attenuation region spanning the frequency range between the two resonant frequencies.

\begin{figure*}[!htbp]
    \centering
    \includegraphics {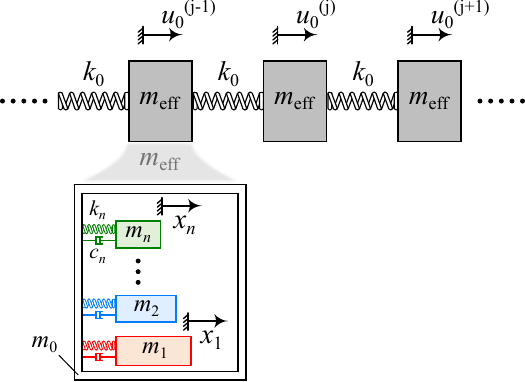}
    \caption{Schematic of the locally resonant mass--spring chain with multi-degree-of-freedom unit cells. The primary lattice consists of effective masses $m_{\mathrm{eff}}$ coupled through springs of stiffness $k_0$, with nearest-neighbor interactions and nodal displacements $u_0^{(j)}$. Each effective mass represents a substructure containing multiple internal resonators, as illustrated in the inset. The inset shows a hierarchical arrangement of internal masses $m_i$ ($i=1,\dots,n$), each connected via springs $k_i$ and dampers $c_i$, forming a multi-resonator system attached to the main mass $m_0$. The internal coordinates $x_i$ describe the relative motion of each resonator.}
    \label{fig:fig1}
\end{figure*}
% The key idea is to tune the mass and frequency ratios of two or more internal resonators such that their individual bandgaps are positioned sufficiently close in frequency to interact. However, the coupling between resonators and the shared host mass reintroduces a passband between adjacent bandgaps even when their frequency ranges nominally overlap, and it is the damping of the resonators that ultimately eliminates this intermediate passband by sustaining a nonzero imaginary part of the effective mass across the bridging frequency region~\cite{meng2023theoretical}. In practice, frequency tuning and damping act together: the former reduces the frequency separation that damping must bridge, and the latter converts the residual positive effective mass region into a continuous attenuation zone, yielding a merged stop band that spans the combined frequency range of the individual resonators.

To make these ideas precise, consider the discrete mass--spring--dashpot model illustrated in Figure ~\ref{fig:fig1}, which represents a 1D chain of locally resonant unit cells. $x_i$ denotes the displacement of the $i^{\mathrm{th}}$ degree of freedom within a representative unit cell, while $u_0^{(j)}$ denotes the host mass displacement of the $j^{\mathrm{th}}$ cell along the chain. Within each unit cell, a host mass $m_0$ carries $n$ parallel internal resonators of mass $m_i$, each connected to the host by a spring of stiffness $k_i$ and a dashpot of coefficient $c_i$. Adjacent unit cells are coupled through an inter-cell spring of stiffness $k_0$. Adopting the complex constant modulus model to represent viscoelastic damping, the dashpot coefficient is expressed as $c_i = k_i \eta_i / \omega$, where $\eta_i$ is the loss factor of the $i^{\mathrm{th}}$ resonator and $\omega$ is the excitation frequency. Under this assumption the system's internal dynamics reduce to a single degree of freedom through the concept of effective mass \cite{meng2023theoretical}, defined as the frequency dependent mass that governs the motion of the host:
\begin{equation}
    \frac{m_\text{eff}}{m_0} = 1 + \sum_{i=1}^{n} \mu_i \,
    \frac{\left(1 - \Omega^2/\Omega_i^2 + \eta_i^2\right) 
    - \mathrm{i}\,\eta_i\left(\Omega^2/\Omega_i^2\right)}
    {\left(1 - \Omega^2/\Omega_i^2\right)^2 + \eta_i^2}
    \label{eq:meff}
\end{equation}
where $\mu_i = m_i/m_0$ is the mass ratio of the $i^{\mathrm{th}}$ resonator, $\Omega = \omega/\omega_1$ is the dimensionless frequency normalized to the natural frequency of the first resonator $\omega_1 = \sqrt{k_1/m_1}$, and $\Omega_i = \omega_i/\omega_1$ with $\omega_i =\sqrt{k_i/m_i}$. The single-resonator result is recovered as the special case $n=1$, for which the effective mass becomes negative over the band $\Omega \in [1,\, \sqrt{1+\mu_1}]$, corresponding to the classical locally resonant bandgap. For $n > 1$, each resonator contributes an additional negative-mass region, but the coupling between resonators and the host modifies the bounds of each region, so the total negative-mass range is not simply the union of the individual single-resonator bands. The physical consequence of this coupling is most clearly seen when damping is absent, $\eta_i = 0$. In that case the imaginary part of Eq.~\eqref{eq:meff} vanishes, and the effective mass is real-valued and singular at each $\Omega_i$. A positive effective mass region, corresponding to a passband, is re-established between the resonances regardless of how closely the individual bandgaps are spaced. When damping is introduced, $\eta_i \neq 0$, the imaginary part of $m_\text{eff}$ becomes nonzero and acts to smooth the singularities. As $\eta_i$ increases, the positive effective mass region between adjacent resonances is progressively suppressed, and the separate bandgaps merge into a single continuous stop band. This damping-driven coalescence is the joining mechanism, and it entails an inherent trade-off: larger $\eta_i$ widens the joined bandgap but simultaneously reduces the peak attenuation depth within it.
 
To quantify the joining mechanism, the propagation of harmonic waves along the chain is examined. The equation of motion of the $j^{\mathrm{th}}$ host mass, after the internal resonator degrees of freedom have been condensed into the effective mass, is:
\begin{equation}
    m_\text{eff}\,\ddot{u}_0^{(j)} + k_0\left(2u_0^{(j)} 
    - u_0^{(j-1)} - u_0^{(j+1)}\right) = 0
    \label{eq:eom_chain}
\end{equation}
where $u_0^{(j)}$ is the host mass displacement of the $j^{\mathrm{th}}$ unit cell and $k_0$ is the inter-cell stiffness. Substituting the Bloch wave ansatz $u_0^{(j)} = U_0\,e^{\mathrm{i}(j\tilde{\kappa} - \omega t)}$, where $\tilde{\kappa} = \kappa a$ is the dimensionless Bloch wave number and $a$ is the lattice spacing, into Eq.~\eqref{eq:eom_chain} yields the dispersion relation:
\begin{equation}
    \cos\tilde{\kappa} = 1 - \frac{m_\text{eff}}{m_0} \cdot 
    \frac{\Omega^2}{2\kappa_0}
    \label{eq:dispersion}
\end{equation}
where $\kappa_0 = k_0/(m_0\omega_1^2)$ is the dimensionless inter-cell stiffness. Since $m_\text{eff}$ is complex-valued when $\eta_i \neq 0$, the Bloch wave number $\tilde{\kappa}$ is also complex, and its imaginary part $\text{Im}(\tilde{\kappa})$ governs the spatial decay of the wave amplitude along the chain. It is important to note that the dispersion relation in Eq.~\eqref{eq:dispersion} is derived under the assumption of an infinite periodic medium. For such a system, the amplitude transmission ratio across $N$ unit cells is obtained directly from the Bloch decay rate as:
\begin{equation}
    \text{TR}_{\infty} = e^{-N\,\text{Im}(\tilde{\kappa})}
    \label{eq:TR_bloch}
\end{equation}
which takes values in $[0,\,1]$, with $\text{TR}_{\infty} = 0$ corresponding to complete attenuation and $\text{TR}_{\infty} = 1$ to perfect transmission. Although $N$ appears in Eq.~\eqref{eq:TR_bloch}, it acts as a scaling parameter on the infinite-medium decay rate rather than a true structural parameter: end effects, boundary reflections, and the distinct behavior of the terminal cells are all neglected under this assumption. To obtain the physically response of a \emph{finite} chain of exactly $N$ unit cells, a back-stepping recursion is employed. Starting from the equation of motion of the last cell, which is connected to only one neighbor, the ratio of displacements between adjacent cells $T_j = u_0^{(j)}/u_0^{(j-1)}$ is defined. For the terminal cell $j = N$, the equation of motion reduces to:
\begin{equation}
    -\Omega^2\,\hat{m}_\text{eff}\,u_0^{(N)} + 
    \kappa_0\left(u_0^{(N)} - u_0^{(N - 1)}\right) = 0,
    \label{eq:eom_last}
\end{equation}
where $\hat{m}_\text{eff} = m_\text{eff}/m_0$ is the normalized effective mass. Dividing through by $u_0^{(N-1)}$ yields the displacement ratio at the terminal cell:
\begin{equation}
    T_{N} = \frac{\kappa_0}{\kappa_0 - 
    \hat{m}_\text{eff}\,\Omega^2}.
    \label{eq:T_last}
\end{equation}
For all interior cells $j = 1, 2, \ldots, N - 1$, the equation of motion is:
\begin{equation}
    -\Omega^2\,\hat{m}_\text{eff}\,u_0^{(j)} + 
    \kappa_0\left(2u_0^{(j)} - u_0^{(j-1)} - 
    u_0^{(j+1)}\right) = 0,
    \label{eq:eom_interior}
\end{equation}
which, upon substituting $T_{j+1}$ from the previously computed step, gives the back-stepping recursion:
\begin{equation}
    T_j = \frac{\kappa_0}{\kappa_0\left(2 - T_{j+1}\right) - 
    \hat{m}_\text{eff}\,\Omega^2}, 
    \quad j = N - 1,\, N - 2,\, \ldots,\, 1.
    \label{eq:recursion}
\end{equation}
The overall finite-chain transmission ratio is then:
\begin{equation}
    \text{TR}_{N} = \left|\frac{u_0^{(N)}}{u_0^{(0)}}\right| = 
    \left|\prod_{j=1}^{N} T_j\right|,
    \label{eq:TR_finite}
\end{equation}
\begin{figure*}[!htbp]
    \centering
    \includegraphics {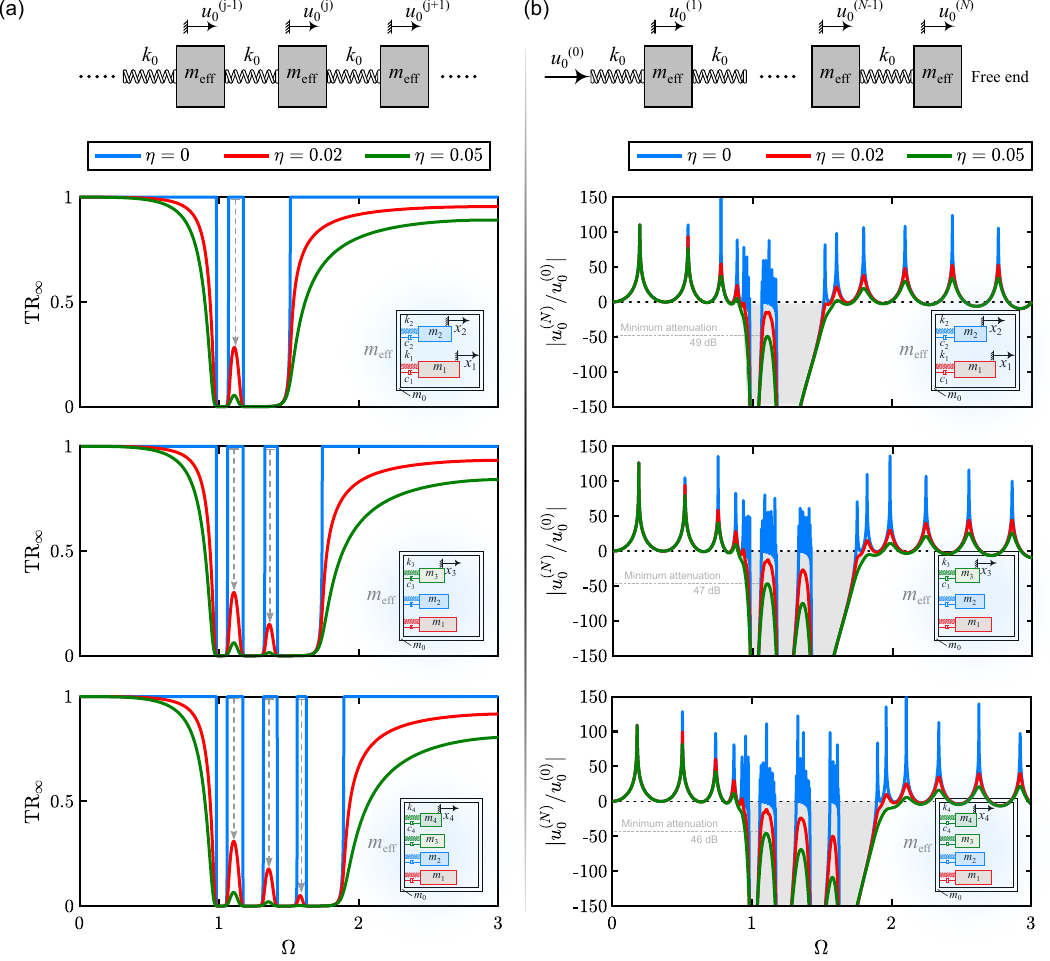}
    \caption{Joining of locally resonant bandgaps in a finite chain. (a) Infinite-medium prediction: The left column shows the Bloch-based transmission ratio $\text{TR}_\infty$ as a function of normalized frequency $\Omega$. Each row corresponds to an increasing number of internal resonators per unit cell (top to bottom), illustrating how multiple locally resonant bandgaps emerge and interact. In the undamped case ($\eta=0$), distinct bandgaps appear around each $\Omega_i$, separated by narrow passbands where $\text{TR}\approx1$. As damping increases, these passbands are progressively suppressed, leading to the merging, “joining”, of adjacent bandgaps into a single broadband attenuation region spanning approximately $\Omega_1$ to $\Omega_n$. (b) Finite-chain response: The right column presents the corresponding frequency response of a finite chain computed via back-stepping recursion. Unlike the smooth Bloch prediction, the finite system exhibits discrete resonance peaks within the passbands and near band edges. For $\eta=0$, sharp transmission peaks appear between bandgaps, consistent with the passbands in (a). With increasing damping, these peaks are strongly attenuated and eventually vanish, while the overall stop-band region becomes continuous.} 
    \label{fig:fig2}
\end{figure*}
which represents the true amplitude attenuation of a finite structure and naturally accounts for end effects through the modified boundary condition at the terminal cell, equation~\eqref{eq:T_last}. The two 
estimates $\text{TR}_{\infty}$ and $\text{TR}_{N}$ converge as $N \to \infty$, confirming that end effects become negligible for sufficiently long chains. For the moderate chain length $N = 10$ used here, the distinction between the two is physically meaningful and is retained deliberately to contrast the infinite-medium Bloch prediction with the finite structural response.

Figure~\ref{fig:fig2} presents the joining phenomenon for $n = 2$, $n = 3$, and $n = 4$ resonators. The system consists of periodically arranged effective masses $m_{\mathrm{eff}}$ coupled by springs of stiffness $\kappa_0=3$, with each unit cell containing multiple internal resonators characterized by mass ratios $\mu_i = [0.4,\,0.4,\,0.2,\,0.1]$ and normalized resonance frequencies $\Omega_i = [1.0,\,1.2,\,1.44,\,1.64]$. Three damping levels are considered: $\eta = 0$ (blue), $0.02$ (red), and $0.05$ (green). The left column shows the Bloch-based transmission ratio $\text{TR}_{\infty}$ on a linear scale from 0 to 1, representing the infinite-medium attenuation prediction. The right column shows the finite-chain frequency response computed via the back-stepping recursion of equations~\eqref{eq:T_last}--\eqref{eq:TR_finite}, which captures the resonance peaks and boundary effects of a real finite structure. For $\eta = 0$, a passband is clearly visible between the individual bandgaps in both columns, appearing as a plateau near $\text{TR}_{\infty} = 1$ in the left column and as a sharp resonance peak in the right column. As $\eta$ increases from $0$ to $0.05$, the passband is progressively suppressed in both representations, and the individual bandgaps coalesce into a single continuous stop band spanning the full frequency range from $\Omega_1 = 1$ to $\Omega_n$. The same qualitative behavior is observed across all three rows, demonstrating that the joining mechanism is general and independent of the number of resonators, while the right column additionally reveals that the finite-chain resonance peaks within the passband are more sensitive to damping than the bulk attenuation within the bandgaps themselves. The results demonstrate that damping plays a critical role in enabling bandgap joining, and that finite-chain resonances are more sensitive to damping than the bulk attenuation predicted by the infinite-medium model. 

The analysis establishes a blueprint for broadband vibration suppression. The remainder of the paper translates this theoretical principle into the design of practical railway vibration absorbers.

\begin{figure*}[!htbp]
    \centering
    \includegraphics {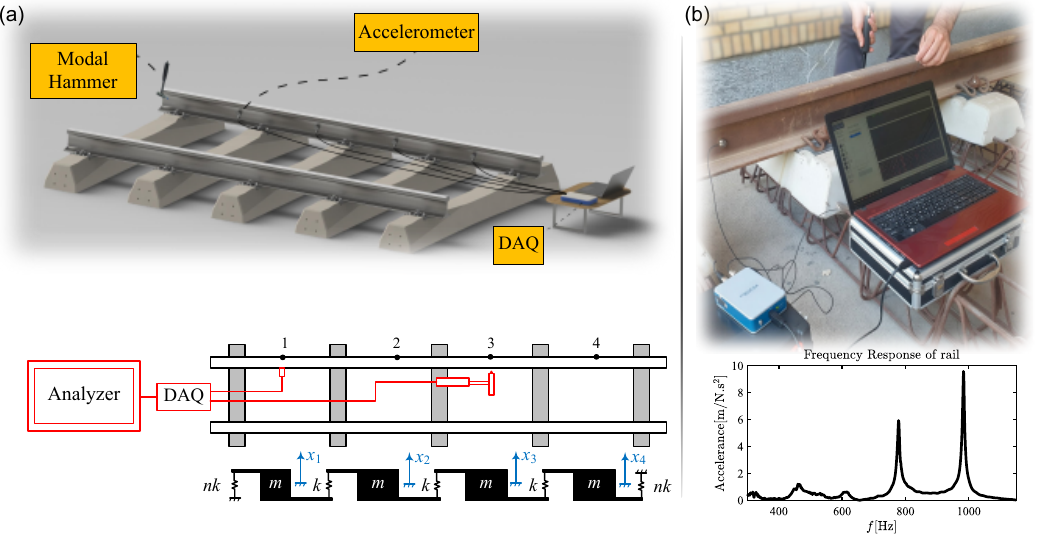}
    \caption{Experimental modal analysis of the railway track and corresponding lumped-parameter representation. (a) Schematic of the impact hammer test used to identify the dynamic characteristics of the UIC60/60E1 rail together with the equivalent four-degree-of-freedom lumped model employed for modal updating. (b) Photograph of the experimental setup and the measured frequency response function (FRF), revealing the dominant flexural resonances used later to design the resonant vibration absorbers.} 
    \label{fig:fig3}
\end{figure*}
\section{Design of Broadband Railway Vibration Absorbers}
The analytical framework developed in the previous section establishes the fundamental design principle for achieving broadband vibration attenuation through the combination of multiple locally resonant bandgaps. This section translates that concept into the design of practical TMDs for railway applications. The design procedure begins with an experimental modal analysis of the rail to identify its dominant vibration modes and develop an equivalent dynamic model suitable for absorber design.
\subsection{Experimental characterization of UIC60 Rail}
To investigate the vibrational behavior of the UIC60/60E1 rail, experimental modal analysis was conducted to characterize its dynamic response under vibration testing. The test configuration is schematically shown in Figure \ref{fig:fig3}(a). The accelerometer data was captured using NI 9230 with a sampling rate of 12800 samples per second, which is much higher than the natural frequencies of the rail. Also, the LabView software was used to capture and record the data, and vibration data analysis was performed using post-processing modal analysis software. Figure \ref{fig:fig3}(b) shows a view of the test along with the frequncy response function of the rail.

Using post-processing software, values of the imaginary part of the accelerance for the first four natural frequencies of the rail are obtained and presented in Table \ref{table1}. 

\begin{table}[!htbp]
  \centering
  \caption{Experimental results for imaginary term of the accelerance}
  \label{tab:accelerance_results}
  \begin{tabular}{@{}c@{\hspace{0.75cm}}c@{\hspace{0.75cm}}c@{\hspace{0.75cm}}c@{\hspace{0.75cm}}c@{}}
    \toprule
    % Header Row
    \begin{tabular}{@{}c@{}}Frequency $\rightarrow$ \\ Accelerance $\downarrow$\end{tabular} &
    $\omega_1$=460 Hz &
    $\omega_2$=604 Hz &
    $\omega_3$=771 Hz &
    $\omega_4$=980 Hz \\
    \midrule
    % Data Rows
    $A_{11}$ & +0.0398 & +0.0489 & +0.1150 & +0.2520 \\
    $A_{12}$ & -0.0148 & -0.0476 & +0.0143 & +0.2450 \\
    $A_{13}$ & +0.0368 & -0.0370 & -0.0753 & +0.2330 \\
    $A_{14}$ & -0.0385 & +0.0606 & -0.0787 & +0.0656 \\
    \bottomrule
  \end{tabular}
  \label{table1}
\end{table}

As shown in Figure \ref{fig:fig3}(a), since the lumped model presented is symmetric with respect to its centerline, the characteristic matrix can be decoupled into two independent \(2\times2\) eigenvalue problems corresponding to the symmetric and antisymmetric vibration modes. Solving these reduced characteristic equations leads to the analytical expressions for the natural frequencies:
\begin{align}
    \omega_{1,3} &= \sqrt{\frac{K}{M}}\sqrt{\frac{(2 + n \pm \sqrt{n^2 + 4})}{2}} 
    \label{eq1}
    \\
    \omega_{2,4} &= \sqrt{\frac{K}{M}}\sqrt{\frac{(4 + n \pm \sqrt{n^2 - 4n + 8})}{2}}
    \label{eq2}
\end{align}

where $n$ is the boundary stiffness ratio coefficient. To find the coefficient $n$, the summation of the relative error is considered as the error parameter, and it is defined as follows:

\begin{equation}
    \text{Error} = \sum_{i=1}^{4} \left( \frac{\left| \omega_i^{\text{Experiment}} - \omega_i \right|}{\omega_i^{\text{Experiment}}} \right)
    \label{eq3}
\end{equation}

\begin{figure*}[!htbp]
    \centering
    \includegraphics {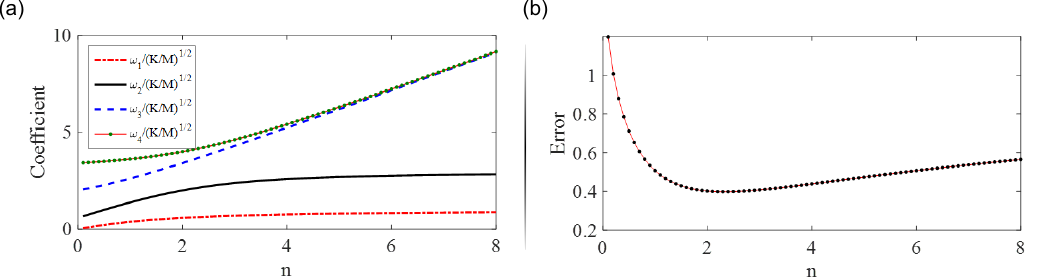}
    \caption{Calibration of the equivalent lumped model through modal identification. (a) Variation of the predicted natural frequencies with the tuning parameter $n$, illustrating its influence on the equivalent mass and stiffness distribution. (b) Total prediction error between the analytical model and the experimentally identified natural frequencies, showing the optimum value of $n$ selected for the modal updating procedure.} 
    \label{fig:fig4}
\end{figure*}

The parameter $n$ modifies the relative stiffness distribution of the simplified model to better represent the dynamic characteristics of the continuous rail. For a prescribed value of $n$, the eigenvalue problem of the lumped system can be solved analytically, yielding the natural frequencies expressed in equations \ref{eq1} and \ref{eq2}. The values of $M$ and $K$ are subsequently determined by matching the analytical frequencies to the experimentally identified modal frequencies, while the optimum value of $n$ is obtained by minimizing the cumulative relative error defined in equation \ref{eq3}. The variation of the coefficient of $\sqrt{K/M}$ in equations \ref{eq1} and \ref{eq2} with changing the coefficient $n$ is shown in Figure \ref{fig:fig4}(a). In addition, effect of the change in coefficient $n$ on the error is depicted in Figure \ref{fig:fig4}(b), where it is shown that the lowest error occurs at $n=2.3$ ($M=41.5$\,kg and $K=349978$\,kN/m).

After doing the modal updating procedure, which is presented in Appendix A, the mass, stiffness, and damping matrices are obtained as follows:
\begin{equation*}
    \mathbf{M} =
    \begin{bmatrix}
        62.6 & 0 & 0 & 0 \\
        0 & 17.57 & 0 & 0 \\
        0 & 0 & 13.06 & 0 \\
        0 & 0 & 0 & 62.6
    \end{bmatrix}
    \, \text{kg}
\end{equation*}
\begin{equation*}
    \mathbf{K} =
    \begin{bmatrix}
        1154927 & -349978 & 0 & 0 \\
        -349978 & 482998 & -133020 & 0 \\
        0 & -133020 & 293657 & -160636 \\
        0 & 0 & -160636 & 965585
    \end{bmatrix}
    \times 10^3 \, \frac{\text{N}}{\text{m}}
\end{equation*}
\begin{equation*}
    \mathbf{C} =
    \begin{bmatrix}
        4493 & -714 & 0 & 0 \\
        -714 & 2295 & -477 & 0 \\
        0 & -477 & 1192 & -276 \\
        0 & 0 & -276 & 4055
    \end{bmatrix}
    \, \frac{\text{N.s}}{\text{m}}
\end{equation*}

Note that the damping matrix is using the Riley damping concept. Furthermore, the modal matrix of the system can be represented as follows:
\begin{equation*}
    \mathbf{\Phi} =
    \begin{bmatrix}
        0.0717 & -0.052 & 0.0788 & 0.0857 \\
        -0.187 & -0.128 & 0.0046 & 0.0762 \\
        0.0947 & -0.148 & -0.115 & -0.064 \\
        -0.033 & -0.032 & 0.071 & -0.06
    \end{bmatrix}
\end{equation*}

The natural frequency values obtained through the EMA procedure are compared with the experimental results in Table \ref{table2}. As shown, the model accurately predicts the system with a maximum error of 2.8\%.

\begin{table}[h!]
    \centering
        \setlength{\tabcolsep}{12pt} %<-- INCREASE THIS VALUE FOR MORE SPACE
    \caption{Comparison of the experimental and modal updated natural frequencies}
    \label{tab:freq_comparison}
    \begin{tabular}{@{}lcccc@{}}
        \toprule
        & $\omega_1$ & $\omega_2$ & $\omega_3$ & $\omega_4$ \\
        \midrule
        Experimental & 460 Hz & 604 Hz & 771 Hz & 980 Hz \\
        Modal Model  & 461 Hz & 587 Hz & 771 Hz & 986 Hz \\
        Error        & 0.2\,\% & 2.8\,\% & 0 & 0.6\,\% \\
        \bottomrule
    \end{tabular}
    \label{table2}
\end{table}

\begin{figure*}[!htbp]
    \centering
    \includegraphics[scale=1]{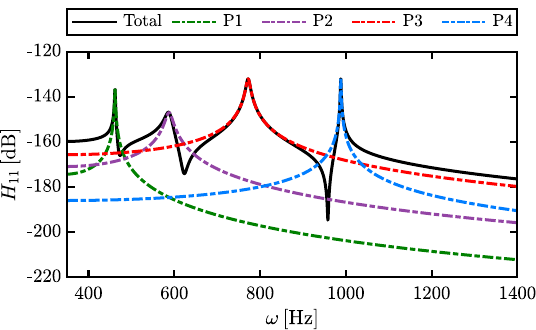}
    \caption{Decomposition of the measured rail receptance into modal contributions. The overall receptance $H_{11}$ is expressed as the superposition of four equivalent single-degree-of-freedom responses, allowing each resonance to be treated independently during the design of the tuned mass dampers.}
    \label{fig:fig5}
\end{figure*}

To design the TMD, the system will be modeled as several separate one-degree-of-freedom systems. This process will be done using the decomposition technique as follows:
\begin{equation}
    H_{ij} = \sum_{r=1}^{4} \frac{\phi_{ir} \times \phi_{jr}}{\omega_r^2 - \omega^2}
\end{equation}
where $H_{ij}$ denotes the receptance (displacement per unit force) between coordinates $i$ and $j$, $\phi_{ir}$ and $\phi_{jr}$ are the modal amplitudes of the $r^{\mathrm{th}}$ vibration mode at the response and excitation locations, respectively, $\omega_r$ is the $r^{\mathrm{th}}$ natural frequency obtained from the updated lumped model, and $\omega$ is the excitation frequency. Since the proposed rail model contains four vibration modes, the receptance is represented as the summation of four modal contributions. As an illustrative example, substituting the identified modal parameters yields the analytical expression for the driving-point receptance $H_{11}$: 
\begin{equation}
    H_{11}=\frac{4.4015 \times 10^{-2}}{8.39 \times 10^6 - \omega^2} + \frac{0.339 \times 10^{-2}}{13.6 \times 10^6 - \omega^2}
    + \frac{19.06 \times 10^{-2}}{23.467 \times 10^6 - \omega^2} + \frac{1.109 \times 10^{-2}}{38.38 \times 10^6 - \omega^2}
\end{equation}
where the receptance $H_{11}$ and its components are depicted in Figure \ref{fig:fig5}. Using this approach, several one-degree-of-freedom equivalent systems are obtained for each degree of freedom.

The equivalent vibratory lumped parameters of the presented system in each natural frequency are shown in Table \ref{table3}.

\begin{table}[h!]
    \centering
    % Adjust the column separation for this table
    \setlength{\tabcolsep}{12pt} %<-- Adjust this value for more/less space
    
    \caption{Specifications of the lumped system in each natural frequency}
    \label{tab:lumped_system}
    \begin{tabular}{@{}lcccc@{}}
        \toprule
        Natural Frequency & $\omega_1$ & $\omega_2$ & $\omega_3$ & $\omega_4$ \\
        \midrule
        $K$ (MN/m)      & 851.14   & 701.64   & 9052.66  & 29853.83 \\
        $M$ (kg)        & 101.44   & 51.58    & 385.75   & 777.82   \\
        $C$ (N.s/m)     & 661.7    & 309.48   & 514.3    & 1606.7   \\
        \bottomrule
    \end{tabular}
    \label{table3}
\end{table}

\subsection{Design methodology of the multi-resonator TMD}
Subsequent to the experimental characterization and modal identification of the rail, the acquired FRF, shown in Figure \ref{fig:fig3}(b), reveals multiple predominant natural frequencies within the 400–1200 Hz range. The resonances at 771 Hz and 980 Hz demonstrate markedly greater amplitudes than adjacent modes, signifying their essential role in structural vibration and noise emission. Furthermore, these frequencies lie within the audible spectrum of human hearing, thus posing a substantial issue regarding vibration-induced acoustic discomfort in subway systems. The main aim of this study is to reduce the vibration response within this particular frequency range. Following the establishment of the theoretical rail model through experimental investigations and the demonstration of bandgap behavior in locally resonant systems, the subsequent step is to devise a realistic and practically implementable TMD configuration that aligns with the actual rail geometry. This section presents a metamaterial-inspired design methodology for dynamic vibration absorbers, wherein multiple locally resonant absorbers are meticulously engineered to concurrently address several predominant frequencies. In contrast to traditional single-frequency absorbers, the proposed method utilizes the principle of distributed local resonances to attain broadband vibration suppression within the critical frequency range determined from the rail frequency response function.

In contrast to the idealized absorber models typically discussed in theoretical analyses, the proposed design must adhere to various practical constraints, such as compatibility with rail geometry, ease of installation, structural integrity, and tunability across multiple resonance frequencies. A modular TMD system, inspired by locally resonant metamaterial concepts, is proposed to tackle these challenges. The aim is to not only eliminate a specific resonant frequency but also to establish multiple local resonances that can mitigate vibrations across a wider frequency spectrum encompassing the primary rail modes. This method integrates the functionality of traditional TMDs with the broadband attenuation properties of acoustic and elastic metamaterials. Subsequently, we will first elucidate the design procedure of the multi-resonator host system, followed by the design of the resonator within the host.

\begin{figure*}[!htbp]
    \centering
    \includegraphics[scale=1]{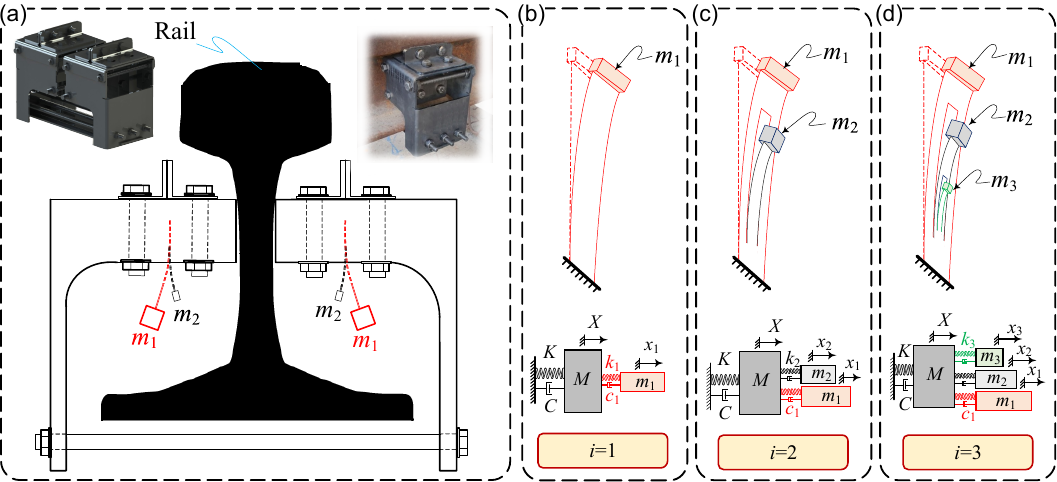}
    \caption{Modular design concept of the proposed multi-resonator TMD. (a) Mechanical configuration of the host structure showing the clamping mechanism, cantilever beams, and interchangeable tip masses used to realize locally resonant absorbers. (b–d) Equivalent dynamic models representing TMDs containing one, two, and three internal resonators, respectively. Although the framework is applicable to an arbitrary number of resonators, only the one- and two-resonator configurations are experimentally investigated in this study because of manufacturing simplicity.}
    \label{fig:fig6}
\end{figure*}

One key aspects of designing the TMDs is ensuring they can be easily attached to the rail. Each TMD unit is designed as a pair of absorbers on both sides of the rail beam, which not only enhances momentum transfer from the rail to the absorber but also allows installation of the absorber without damage to the rail. To do this, each vibration absorber unit should be clamped on both sides of the rail beam, with the clamp jaws using screws. The system consists of two vibrating absorbers and a box housing the vibrating components. Two cantilever beams, which directly interact with the tip mass, are made of spring steel (AISI 1095) with a thickness of 1 mm and function as the spring ($k$) component of the system. The spring coefficient of these beams can be calculated using the equation $k=3EI/L^3$, where $E$ denotes for young's modulus, $I$ area moment of inertia and $L$ is the length of spring beam\cite{rao2019vibration}. In this system, the tip mass of the beam, which acts as the TMD mass, consists of a combination of 2, 4, or 6 masses, allowing for tuning and better performance in different applications, as multi-layer cantilever beams can act better in controlling passive vibrations \cite{Hodaei2024}. Figure \ref{fig:fig6}(a) demostrates two symmetrical sections make up the vibration absorber system, which is intended to reduce lateral vibrations transmitted through the rail. The tip masses and spring steel cantilever beams are the main functional elements, dissipating vibrational energy. Because the system is fully bolted together, it is modular and straightforward to modify. The absorber is fastened to the rail by the holding components, which include the base and supporting brackets and act as the structural foundation. The cantilever beams oscillate due to the rail's vibrations, which causes the attached tip masses to respond dynamically. The tip masses help to adjust the system's inherent frequency, and the spring steel beams flex, storing and releasing energy in a controlled way. Because of this combination, undesired vibrations can be efficiently absorbed and dissipated, lessening their effect on the main structure. 

Considering the primary single-degree-of-freedom system on Figure \ref{fig:fig6}(b), (c), and (d), where characterized by its mass $M$, stiffness $K$, and damping coefficient $C$ subjected to external harmonic excitations, a locally resonator absorber can be defined by its mass $m_1$, stiffness $k_1$, and damping coefficient $c_1$, is attached to the main structure to suppress its vibration amplitude at resonance. For a system with with $i+1$ degrees of freedom which means hosting $i$ absorbers, as described in Figure \ref{fig:fig6}, the coupled equations of motion are:

\begin{equation}
    \textbf{M}\ddot{x}+\textbf{C}\dot{x}+\textbf{K}x=f
\end{equation}
where $\textbf{M}$, $\textbf{C}$, and $\textbf{K}$ are the global mass, damping, and stiffness matrices, given by: 
\begin{equation}
\begin{split}
\mathbf{M} &= 
\begin{bmatrix}
M & 0 & \cdots & 0 \\
0 & m_1 & \cdots & 0 \\
\vdots & \vdots & \ddots & \vdots \\
0 & 0 & \cdots & m_n
\end{bmatrix} \\[1ex]
\mathbf{C} &= 
\begin{bmatrix}
C + \sum c_i & -c_1 & \cdots & -c_n \\
-c_1 & c_1 & \cdots & 0 \\
\vdots & & \ddots & \vdots \\
-c_n & 0 & \cdots & c_n
\end{bmatrix} \\[1ex]
\mathbf{K} &= 
\begin{bmatrix}
K + \sum k_i & -k_1 & \cdots & -k_n \\
-k_1 & k_1 & \cdots & 0 \\
\vdots & & \ddots & \vdots \\
-k_n & 0 & \cdots & k_n
\end{bmatrix}
\end{split}
\end{equation}

The FRF between an excitation force at the first degree of freedom and its corresponding displacement response is given as:
\begin{equation}
    H_{11}=\zeta_1^T [-\omega^2 \textbf{M}-i\omega\textbf{C}+\textbf{K}]_{11}^{-1}\zeta_1
    \label{H_11}
\end{equation}
where $\zeta_1$ is the unit vector for the first DOF. The magnitude $|H_{11}|$ represents the vibration transmissibility and is used as a measure of vibration suppression performance. An optimization process is designed to lower the highest value of the receptance frequency response function $H_{11}$ of the main degree of freedom when harmonic excitation is present. For a specific number of connected subsystems, the global mass, damping, and stiffness matrices are organized in a star-coupled arrangement, with each attachment linked directly to the main mass. The mechanical properties of the attachments follow exponential scaling laws $m_i=M/A_i$, $c_i=C/B_i$, and $k_i=K/G_i$ with design variables $A$, $B$, and $G$ constrained to be larger than $1.5$ to ensure physically meaningful and monotonically decreasing distributions. The frequency response is obtained by solving the complex dynamic stiffness equation (\ref{H_11}) over a prescribed frequency band, and the receptance $H_{11}$ is extracted as the displacement of the primary mass due to a unit force applied at the same location. For every value of $i$ (from $1$ to $6$), a nonlinear constrained optimization problem is established, wherein the objective function represents the maximum magnitude of $|H_{11}|$ across the design frequency range. This peak value is the worst-case vibration amplification, so it is a good measure of how well vibration suppression works. A gradient-based sequential quadratic programming algorithm is used to do the optimization. It can also be combined with a multistate strategy to make it less sensitive to local minima. The best sets of ($A_i$, $B_i$, $G_i$) give the best distributions of mass, damping, and stiffness among the attachments. Theses values are shown in Table \ref{optimization}.
\begin{table}[htbp]
\centering
 \setlength{\tabcolsep}{12pt} 
\caption{Optimum values for $A_i$, $B_i$ and $G_i$ and maximum receptance}
\label{tab:optimum_values_wide}
\begin{tabular}{cccccc}
\toprule
$i$ & $A_i$ & $B_i$ & $G_i$ & $\lvert H_{11} \rvert_{\max}$ (mm/N) & Decrement (\%) \\
\midrule
0 & -     & -   & -     & 0.447 & -    \\
1 & 4.105 & 1.5 & 6.123 & 0.169 & 62.1 \\
2 & 3.693 & 1.5 & 6.095 & 0.153 & 65.8 \\
3 & 1.581 & 1.5 & 2.608 & 0.130 & 70.9 \\
4 & 2.013 & 1.5 & 3.530 & 0.125 & 72.0 \\
5 & 3.587 & 1.5 & 6.153 & 0.149 & 66.7 \\
6 & 1.554 & 1.5 & 2.576 & 0.110 & 75.4 \\
\bottomrule
\label{optimization}
\end{tabular}
\end{table}

It is noteworthy that the optimum values of ($B_i$) in Table 4 are equal to the imposed lower bound for all investigated absorber configurations. This indicates that, for the present objective function—namely minimization of the peak receptance of the primary system over the target frequency range—the optimization consistently favors increasing the absorber damping up to the maximum level permitted within the admissible design space. In other words, the damping-related optimum is constraint-active rather than interior. This behavior does not imply that ($B_i=1.5$) is a universal physical optimum; rather, it shows that within the practical range considered here, stronger absorber damping improves suppression of the dominant resonance peaks. The selected bound was retained because it corresponds to the highest absorber damping level judged to remain compatible with the intended cantilever-type resonator implementation and with preservation of a clear locally resonant absorber response.

\begin{figure*}[!t]
    \centering
    \includegraphics[width=\textwidth]{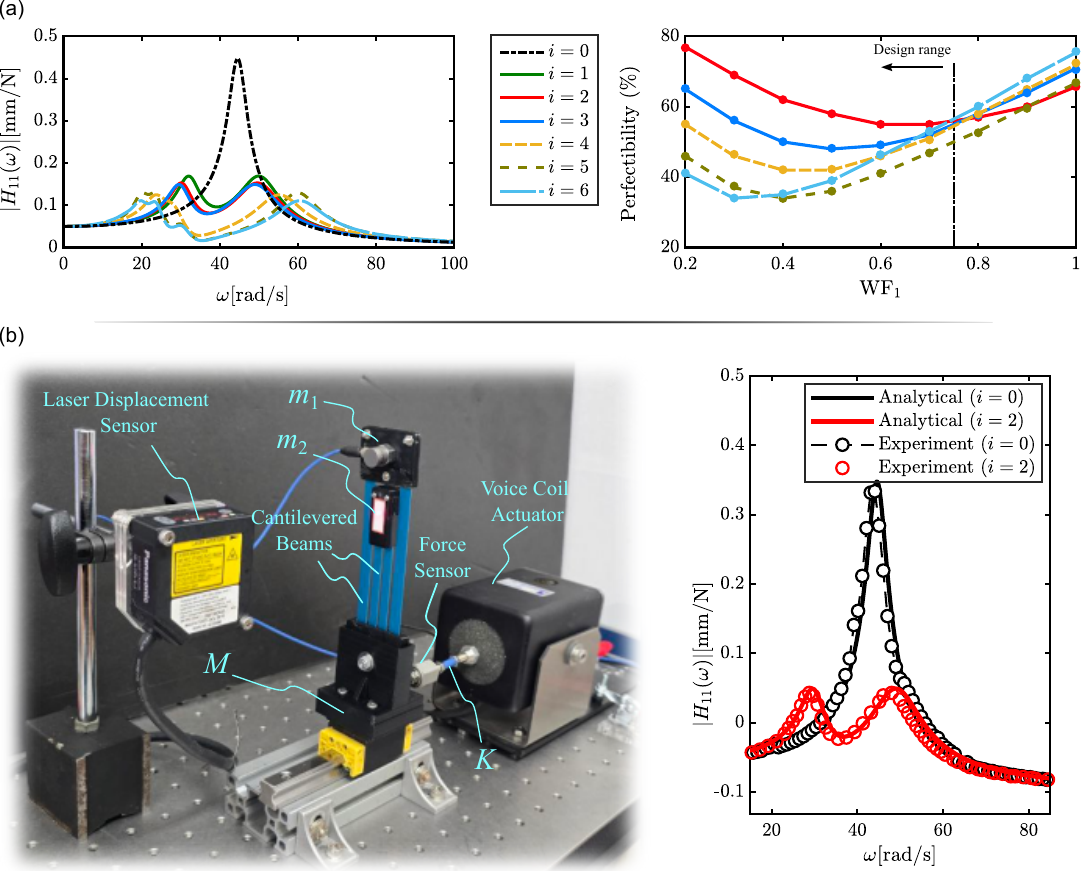}
    \caption{Numerical optimization and experimental validation of the proposed multi-resonator design methodology. (a) Numerical optimization results based on the optimum values of the mass, damping, and stiffness scaling parameters ($A_i$, $B_i$, and $G_i$). The left panel compares the receptance of the primary system equipped with $i=1-6$ dynamic vibration absorbers (TMDs) against the baseline system without a TMD ($i=0$). The right panel illustrates the variation of the proposed perfectibility metric with the weighting factor $WF_1$, highlighting the design range and the optimum TMD configuration. (b) Experimental validation of the bi-resonator ($i=2$) configuration. The left panel shows the laboratory setup used to evaluate the proposed resonator, including the host structure, cantilever beams, excitation system, and measurement instruments. The right panel compares the analytical and experimental receptance of the systems without a TMD ($i=0$) and with the optimized bi-resonator configuration ($i=2$), demonstrating excellent agreement and validating the proposed design methodology.}
    \label{fig:fig7}
\end{figure*}

Left portion of Figure \ref{fig:fig7}(a) shows the frequency response function of the rail system with and without TMDs and with different numbers of TMDs attached. As you can see, adding more TMDs usually makes resonance peaks less noticeable and makes vibration suppression better for each configuration of the attached subsystems. But making $i$ bigger makes the manufacturing process more complicated and makes the design less practical. So, when designing the whole system, both vibration reduction performance and how well it works in real life must be taken into account at the same time. To find this balance, a Perfectibility parameter is added. This is how it is defined:
\begin{equation}
    \text{Perfectibility}=\text{WF}_{1} (1-H_i⁄H^* )+\text{WF}_{2}^i
    \label{Perf}
\end{equation}
where $H_i$ denotes the maximum value of the receptance $H_{11}$ for $i^\text{th}$ configuration, $H^*$ is the maximum value of the system receptance without TMD, and \text{WF}${_1}$ and \text{WF}${_2}$ are weighting factors associated with vibration suppression performance and the simplicity (practical feasibility) of the TMD system, respectively. Right portion of Figure \ref{fig:fig7}(a) shows how the perfectibility parameter changes with \text{WF}${_1}$. In this study, the weighting factors are limited so that \text{WF}${_1}$+\text{WF}${_2}$$=1$. \text{WF}${_1}$$<0.75$, which is called the `Design Range,' the setup with two TMDs gives the best overall performance, as shown.

\subsection{Experimental validation of the resonator design}

Before evaluating the proposed TMDs on the railway track, the dynamic behavior of the fabricated resonators was experimentally verified to ensure that their measured characteristics agreed with the analytical design. Since the optimization procedure identified the bi-resonator configuration ($i=2$) as the preferred design based on the proposed perfectibility criterion, this configuration was selected for experimental validation. The objective of this experiment is to demonstrate that the designed resonators reproduce the target resonant frequencies and vibration attenuation predicted by the analytical model, thereby validating the proposed design methodology.

A laboratory-scale vibratory system with equivalent properties of $M=0.103$ kg, $K=214$ N/m, and $\zeta=0.056$ was constructed, as illustrated in Figure~\ref{fig:fig7}(b). Harmonic excitation was applied using a voice-coil actuator, while the input force was monitored by a force sensor and the resulting displacement of the primary mass was measured using a laser displacement sensor. The resonator parameters were selected according to the optimum scaling factors reported in Table~9, and the receptance frequency response function (FRF) was obtained experimentally.

Figure~\ref{fig:fig7}(b) compares the analytical and experimental receptance of both the baseline system ($i=0$) and the optimized bi-resonator configuration ($i=2$). Excellent agreement is observed between the predicted and measured responses, with both the resonance frequencies and peak amplitudes accurately captured by the analytical model. These results confirm that the proposed design methodology successfully translates the theoretical resonator parameters into a practical hardware implementation, providing confidence in applying the designed TMDs to the railway track in the following section.

It should be emphasized that the optimization results of Table \ref{optimization} were used primarily to determine the most effective absorber architecture and the corresponding relative distribution of resonator properties, rather than to prescribe a unique set of hardware dimensions. In particular, the optimization showed that increasing the number of attached resonators generally improves attenuation, but at the expense of fabrication complexity and reduced practical feasibility. Based on the proposed perfectibility criterion, the one-resonator and two-resonator configurations were therefore selected as the most suitable candidates for experimental realization, with the bi-resonator configuration identified as the preferred broadband design. The physical TMDs fabricated in the subsequent sections were designed as cantilever-based realizations of these optimized configurations. Because the manufactured absorbers employ discrete tip masses and beam elements, the nondimensional optimization parameters ($A_i,B_i,G_i$) were translated into equivalent target resonator masses and resonance frequencies compatible with the rail-mounted hardware. Accordingly, the single-resonator experiments reported in section~\ref{experiment_section} should be interpreted as a practical tuning study around the optimized one-resonator design, where three realizable mass ratios were tested to identify the best-performing physical implementation. The bi-resonator absorber was then constructed by taking the best-performing single-resonator configuration as the primary resonator and adding a secondary resonator according to the optimized two-resonator design strategy, thereby providing the experimental realization of the broadband absorber concept developed in this section.
 % To validate the results, the configuration employing two DVAs ($i=2$) was experimentally investigated. Accordingly, a vibratory system with $M=0.103$ kg, $K=214$ N/m, and $\zeta=0.056$ was considered, as illustrated in Figure \ref{fig:fig7}(b). In this configuration, the parameters $A_i$ and $B_i$ were selected based on the values reported in Table 9, and the corresponding measured frequency response function is presented.

\section{Experimental Validation on Railway Tracks}
\label{experiment_section}
Following the design and validation of the proposed resonators, this section evaluates their effectiveness when integrated into a practical railway vibration mitigation system. The fabricated TMDs are installed on a 3-m UIC60 rail to experimentally assess their ability to suppress flexural vibrations under impact excitation. Both single-resonator and bi-resonator TMD configurations are investigated to examine the influence of multiple internal resonators on broadband vibration attenuation. Although the proposed design methodology can accommodate an arbitrary number of resonators, only the one- and two-resonator configurations are considered in this study to maintain manufacturing simplicity while demonstrating the practicality and effectiveness of the proposed broadband design concept.

The identical experimental setup previously described in Figure~\ref{fig:fig3} was employed to evaluate the vibration response of the railway track. The FRFs were obtained using impact hammer excitation, while an IEPE accelerometer (GT-AP2037) positioned at Node~1 measured the corresponding vibration response under both TMD-present and TMD-absent conditions. In all experiments, six identical TMD units were periodically installed along the 3-m rail, forming a finite chain of locally resonant absorbers. This configuration serves as a practical realization of the finite resonator array discussed in section~\ref{joining_bandgaps}, allowing the collective vibration attenuation characteristics of the proposed broadband design methodology to be experimentally evaluated.
 
\begin{figure*}[!t]
    \centering
    \includegraphics[scale=0.8]{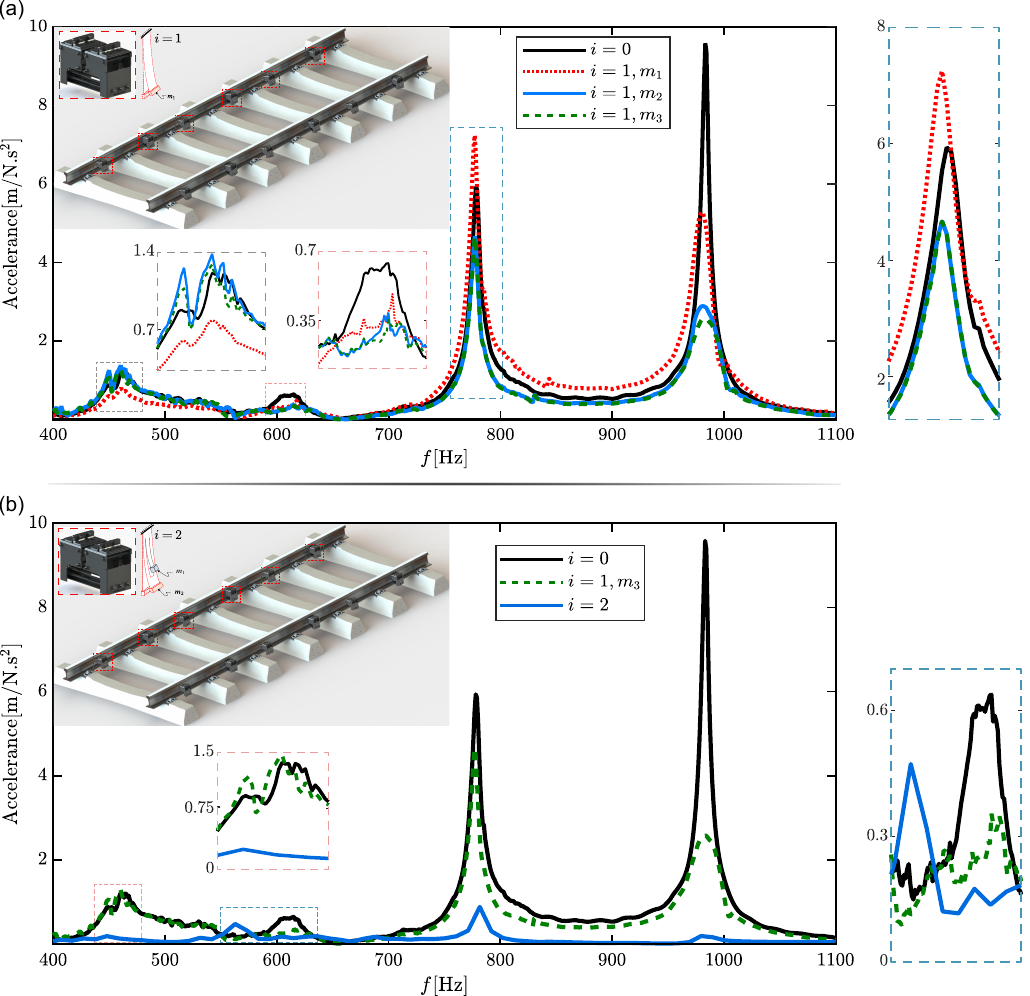}
    \caption{Experimental evaluation of the proposed TMD configurations on a 3-m railway track. The frequency response functions (FRFs) were obtained using impact hammer testing, with the vibration response measured by an IEPE accelerometer (GT-AP2037) positioned at Node 1 under both TMD-present and TMD-absent conditions. Six identical TMD units were periodically installed along the rail to evaluate their vibration suppression performance. (a) Performance of the single-resonator TMD for three different resonator masses. The configuration with resonator mass $m_1$ ($\mu=0.22$) provides effective attenuation at low frequencies but amplifies the third resonance. Increasing the resonator mass to $m_2$ ($\mu=0.44$) shifts the attenuation toward higher frequencies while reducing its effectiveness at low frequencies. The largest resonator mass, $m_3$ ($\mu=0.66$), achieves broadband vibration suppression and effectively reduces the vibration amplitude at all dominant resonance peaks. (b) Performance of the bi-resonator TMD. The first resonator corresponds to the optimum single-resonator design identified in panel (a), while a second resonator with one-third of the primary resonator mass is incorporated to broaden the attenuation bandwidth. The FRFs of the optimized single-resonator and bi-resonator configurations are compared, demonstrating the enhanced broadband vibration suppression achieved by the proposed two-resonator design.}
    \label{fig:8}
\end{figure*}

The first experiment investigates the performance of the single-resonator TMD by varying the resonator mass while maintaining the same finite-chain arrangement. The baseline resonator mass ($m_1$) was selected from the optimized single-resonator design reported in Table~\ref{table4}, where the corresponding scaling parameter ($A_i$) determines the target absorber mass. However, because the practical implementation of the rail-mounted TMD is subject to manufacturing tolerances, attachment uncertainties, and simplifications associated with the equivalent analytical model, two additional realizable tuning cases were also examined by increasing the resonator mass to ($2m_1$) and ($3m_1$), denoted here as ($m_2$) and ($m_3$), respectively. This experimental sweep was therefore introduced to identify the best-performing physical implementation in the neighborhood of the optimized design point rather than to define independent ad hoc absorber configurations. As shown in Figure~\ref{fig:8}(a), the configuration with resonator mass ($m_1$) provides effective vibration suppression at the lower resonant frequencies but amplifies the third resonance. Increasing the resonator mass to ($m_2$) shifts the attenuation toward higher frequencies at the expense of reduced low-frequency performance. The largest resonator mass, ($m_3$), provides the most balanced response, effectively suppressing all dominant resonance peaks and yielding the widest attenuation bandwidth among the three configurations. The quantitative results summarized in Table~\ref{table5} show that this optimum configuration reduces the vibration amplitude by 22\%--74\% at the dominant resonant frequencies, while having only a negligible influence on the first resonance. It should also be noted that the reported TMD performance represents the percentage change in vibration amplitude relative to the baseline case without TMDs, where positive values indicate vibration reduction and negative values indicate vibration amplification.

% The first experiment investigates the performance of the single-resonator TMD by varying the resonator mass while maintaining the same finite-chain arrangement. As shown in Figure~\ref{fig:8}(a), the configuration with resonator mass $m_1$ ($\mu=0.22$) provides effective vibration suppression at the lower resonant frequencies but amplifies the third resonance. Increasing the resonator mass to $m_2$ ($\mu=0.44$) shifts the attenuation toward higher frequencies at the expense of reduced low-frequency performance. The largest resonator mass, $m_3$ ($\mu=0.66$), provides the most balanced response, effectively suppressing all dominant resonance peaks and yielding the widest attenuation bandwidth among the three configurations. The quantitative results summarized in Table~\ref{table5} show that this optimum configuration reduces the vibration amplitude by 22\%--74\% at the dominant resonant frequencies, while having only a negligible influence on the first resonance. It should also be noted that the reported TMD performance represents the percentage change in vibration amplitude relative to the baseline case without TMDs, where positive values indicate vibration reduction and negative values indicate vibration amplification.
The second experiment evaluates the proposed bi-resonator TMD using the same finite-chain installation. The primary resonator corresponds to the optimized single-resonator design identified in the previous experiment, while the secondary resonator was introduced based on the optimized two-resonator configuration reported in Table~\ref{optimization}. Specifically, the mass of the second resonator was selected using the corresponding mass-scaling parameter ($A_2$) obtained from the numerical optimization of Section~3.2, thereby ensuring that the fabricated bi-resonator absorber remained consistent with the optimized broadband design framework. In the manufactured implementation, this resulted in a secondary resonator mass approximately equal to one-third of the primary resonator mass, which was compatible with the bandgap-joining strategy developed in Section~\ref{joining_bandgaps}.

Figure~\ref{fig:8}(b) compares the measured frequency response of the optimized single-resonator and bi-resonator configurations. The addition of the second resonator further broadens the effective attenuation bandwidth and produces a more uniform reduction of the resonance peaks across the frequency range of interest. The quantitative results summarized in Table~\ref{table5} show that the bi-resonator TMD reduces the vibration amplitudes by 86\%, 29\%, 85\%, and 98\% at the first through fourth resonant frequencies, respectively. Compared with the optimized single-resonator design, the proposed bi-resonator configuration substantially improves the attenuation of the first, third, and fourth resonances while maintaining effective suppression of the second resonance. These results experimentally demonstrate that incorporating multiple locally resonant elements within each TMD enhances the broadband vibration suppression capability of the finite resonator array while maintaining a compact and practical absorber design.

% The second experiment evaluates the proposed bi-resonator TMD using the same finite-chain installation. The primary resonator corresponds to the optimum single-resonator design identified in the previous experiment, while a secondary resonator with one-third of the primary resonator mass is incorporated according to the bandgap-joining strategy developed in Section~\ref{joining_bandgaps}. Figure~\ref{fig:8}(b) compares the measured frequency response of the optimized single-resonator and bi-resonator configurations. The addition of the second resonator further broadens the effective attenuation bandwidth and produces a more uniform reduction of the resonance peaks across the frequency range of interest. The quantitative results summarized in Table~\ref{table5} show that the bi-resonator TMD reduces the vibration amplitudes by 86\%, 29\%, 85\%, and 98\% at the first through fourth resonant frequencies, respectively. Compared with the optimized single-resonator design, the proposed bi-resonator configuration substantially improves the attenuation of the first, third, and fourth resonances while maintaining effective suppression of the second resonance. These results experimentally demonstrate that incorporating multiple locally resonant elements within each TMD enhances the broadband vibration suppression capability of the finite resonator array while maintaining a compact and practical absorber design.

\begin{table}[htbp]
    \centering
    % Set the space between columns. Adjust '8pt' or '10pt' for layout fitting.
    \setlength{\tabcolsep}{8pt}

    \caption{Combined experimental results of the system Accelerance without and with TMD configurations}
    \label{tab:dva_combined_results}
    \begin{tabular}{@{}ccccccc@{}}
        \toprule
        \multirow{2}{*}{Parameter} & \multirow{2}{*}{Frequency} & \multirow{2}{*}{without TMD ($i=0$)} & \multicolumn{4}{c}{with TMD} \\
        \cmidrule(lr){4-7}
        & & & $i=1,m_1$ & $i=1,m_2$ & $i=1,m_3$ & $i=2$ \\
        \midrule
        
        % Block for 460 Hz
        $A$ (m/s$^2$) & \multirow{2}{*}{460 Hz} & 1.30 & 0.80 & 1.38 & 1.32 & 0.18 \\
        \cmidrule(l){1-1}
        Performance of TMD & & - & +38 & -6 & -1.5 & +86 \\
        \midrule
        
        % Block for 604 Hz
        $A$ (m/s$^2$) & \multirow{2}{*}{604 Hz} & 0.68 & 0.52 & 0.48 & 0.29 & 0.48 \\
        \cmidrule(l){1-1}
        Performance of TMD & & - & +24 & +29 & +57 & +29 \\
        \midrule
        
        % Block for 771 Hz
        $A$ (m/s$^2$) & \multirow{2}{*}{771 Hz} & 6.05 & 7.20 & 4.72 & 4.70 & 0.88 \\
        \cmidrule(l){1-1}
        Performance of TMD & & - & -19 & +22 & +22 & +85 \\
        \midrule

        % Block for 980 Hz
        $A$ (m/s$^2$) & \multirow{2}{*}{980 Hz} & 9.50 & 5.10 & 2.92 & 2.50 & 0.19 \\
        \cmidrule(l){1-1}
        Performance of TMD & & - & +46 & +69 & +74 & +98 \\
        \bottomrule
        \label{table5}
    \end{tabular}
\end{table}

\section{Random vibration analysis}
The experimental results presented in section~\ref{experiment_section} demonstrate the effectiveness of the proposed TMDs in suppressing rail resonances under controlled impact excitation. In practical railway applications, the excitation acting on the track is stochastic and arises from a combination of wheel--rail interaction mechanisms, including track irregularities, vehicle dynamics, and local contact variations. A rigorous representation of such loading would require a dedicated vehicle--track interaction model together with an irregularity-based excitation spectrum. The objective of the present section is therefore more modest: rather than reproducing the full complexity of operational railway loading, it provides a stochastic assessment of the proposed TMDs under an idealized broadband random excitation. To this end, a band-limited white Gaussian excitation is adopted as a simplified input model to examine how the rail response changes when broadband vibration energy is distributed over the frequency range containing the dominant resonances of the system.

% demonstrate the effectiveness of the proposed TMDs in suppressing the rail resonances under controlled impact excitation. In practical railway applications, however, the excitation acting on the track is inherently random due to wheel--rail interactions, surface irregularities, vehicle dynamics, and other operational uncertainties. Consequently, evaluating the performance of the proposed TMDs under stochastic loading conditions is essential for assessing their effectiveness in realistic service environments. The objective is to quantify the reduction in the statistical vibration response after installing the proposed TMDs by comparing the output power spectral densities and the corresponding root-mean-square (RMS) vibration levels of the rail before and after vibration mitigation. 

For linear dynamic systems, random vibration analysis is conveniently performed in the frequency domain using PSD functions, which characterize how the excitation energy is distributed over frequency. Once the PSD of the input excitation is known, the statistical response of the system can be obtained directly through the frequency response function, enabling efficient prediction of the vibration energy transmitted to the rail. The Fourier transform of the autocorrelation function of a process $x$ yields the process spectral density. The spectral density is calculated using the following formula:

\begin{equation}
    \mathrm{S}_x(\omega) = \frac{1}{2\pi} \int_{-\infty}^{\infty} \mathrm{R}_x(\tau) e^{-i\omega\tau} \,\mathrm{d}\tau
\end{equation}
Inversely, $R_x$ can be written as follows:
\begin{equation}
    \mathrm{R}_x(\tau) = \int_{-\infty}^{\infty} \mathrm{S}_x(\omega) e^{i\omega\tau} \,\mathrm{d}\omega
\end{equation}

% In general, there is no certainty that the system's input forces are correlated. However, this study considers the hypothesis that a white Gaussian noise excites all four system nodes with the same spectral and statistical behavior, taking into account a part of the rail and real conditions. 

In the present study, the external excitation is modeled as a band-limited white Gaussian process applied to the four generalized coordinates of the equivalent rail system. The same input spectral level is assigned to each coordinate in order to provide a uniform broadband excitation benchmark for comparing the uncontrolled and controlled responses of the rail over the frequency range of interest. This assumption does not aim to reproduce the detailed spectral structure of actual wheel--rail forces; rather, it is adopted as a simplified stochastic loading model that enables a first-order assessment of the influence of the proposed TMDs on the rail RMS response. This force and any other type of force represented in the frequency domain can be used to calculate the mean square response using the spectral density method. The square root of the average squared values of the vibration signal over time is used to calculate the RMS amplitude. RMS is the most important factor because it shows the vibration's strength and potential damage to a structure. Using the following equation, the system's mean square response can be calculated using the spectral density of the applied forces on the track:

\begin{equation}
    \mathrm{E}[Z^2] = \int_{-\infty}^{\infty} \left| H(\omega) \right|^2 \mathrm{S}_x(\omega) \,\mathrm{d}\omega
\label{eq9}
\end{equation}
Finding a closed-form solution to the integral will be challenging because of the complicated equations obtained in this study to calculate the shape of the spectral density. However, for a 1DOF damped system subjected to white Gaussian noise input, Newland \cite{newland2006mechanical} presented a closed-form solution to the equation mentioned above. It is given by
\begin{equation}
    \mathrm{E}[Z^2] = \frac{\pi S_0}{kc}
\end{equation}
The white Gaussian noise is defined as having a constant spectral density over the entire frequency range. The mean square value of white noise is logically infinite because the frequency bandwidth in white noise extends to infinity. Therefore, this idea is still purely theoretical and has no actual application. When the bandwidth significantly exceeds the required frequencies, the term band-limited white noise can be used. In this case, the autocorrelation function when the lower frequency limit $\omega_1=0$, becomes \cite{Bakhtiari-Shahri2019}:
\begin{equation}
    \mathrm{R}_x(\tau) = \frac{2S_0 \sin(\omega_2\tau)}{\tau} \\
\end{equation}
Considering the fourth natural frequency of the system (976 Hz), it is assumed that $\omega_2=2000$Hz. So, the spectral density of the band-limited white noise would be:
\begin{equation}
\mathrm{S}_x = \frac{0.785S_0}{\pi} \\
    \end{equation}
    Substituting this value in equation \ref{eq9} results in:
\begin{equation}
    \mathrm{E}[Z_i^2] = \frac{0.785S_0}{k_i c_i}
\end{equation}

Therefore, decomposing the system into four independent single-degree-of-freedom (1-DOF) damped systems significantly simplifies the calculations, as demonstrated by the following equations (with detailed derivation provided in Appendix B).
\begin{align}
\mathrm{S}_{y_r} &= \sum_{r=1}^{4} \sum_{i=1}^{4} \sum_{j=1}^{4} \varphi_{ri} \, \varphi_{rj} \, \mathrm{S}_{z_{ij}}(\omega) 
\label{eq14}\\
\mathrm{E}[y_r^2(t)] &= \sum_{r=1}^{4} \sum_{i=1}^{4} \sum_{j=1}^{4} \varphi_{ri} \, \varphi_{rj} \int_{-\infty}^{\infty} \mathrm{S}_{z_{ij}}(\omega) \, \mathrm{d}\omega
\label{eq15}
\end{align}

\begin{figure}[!t]
    \centering
    \includegraphics[scale=0.85]{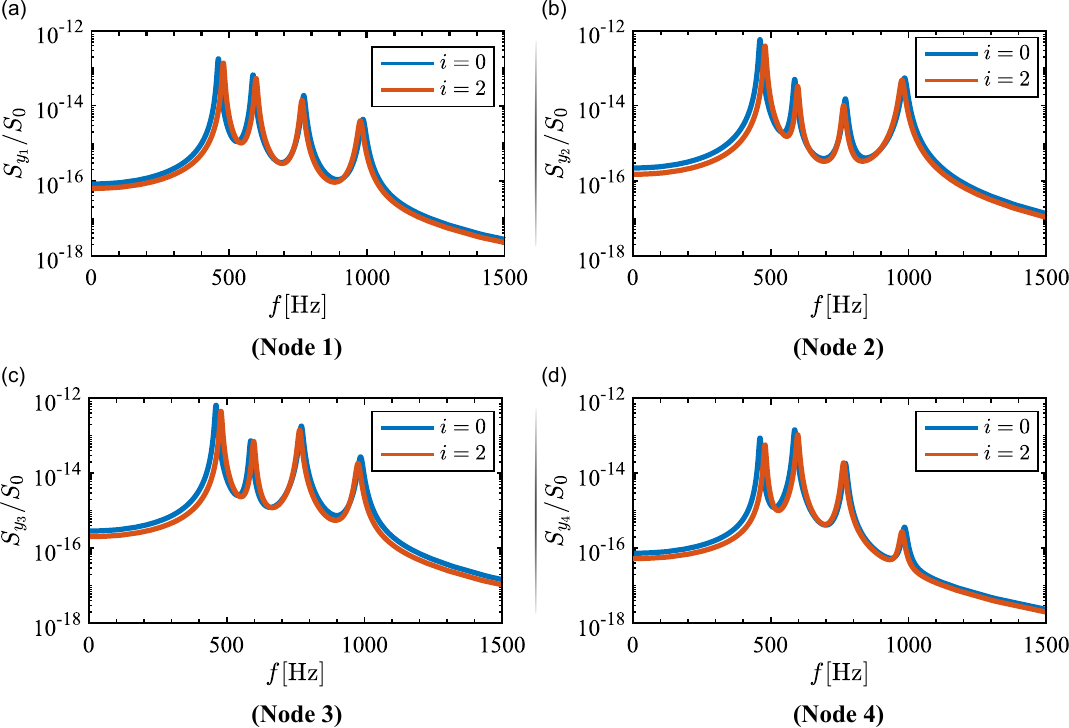}
    \caption{Spectral density without TMD and with bi-resonator TMD in different DOF}
    \label{fig:9}
\end{figure}

After installation of the TMD units, the rail--absorber assembly can in principle be modeled as a higher-order coupled system including the additional resonator degrees of freedom. In the present study, however, a reduced-order representation is adopted for the random-vibration analysis, in which the effect of the attached absorbers is incorporated through equivalent modifications of the rail dynamic properties over the frequency range of interest. This simplification is supported by the relatively small total mass of the installed absorbers compared with the host rail system, as well as by the objective of this section, which is to compare the controlled and uncontrolled rail response using the experimentally validated low-order rail model rather than to reconstruct the full absorber-resolved dynamics. Accordingly, the stochastic analysis should be interpreted as a reduced-order assessment of the influence of the proposed TMDs on the rail response, rather than as a full higher-order simulation of the coupled rail--absorber assembly. Based on the resulting reduced-order formulation, the spectral density functions of the four rail outputs are shown in Figure~\ref{fig:9}.
% After installation of the TMD units, the rail--absorber assembly can in principle be modeled as a higher-order coupled system including the additional resonator degrees of freedom. In the present study, however, a reduced-order representation is adopted for the random-vibration analysis, in which the effect of the attached absorbers is incorporated through equivalent modifications of the rail dynamic properties over the frequency range of interest. This simplification is justified by the relatively small total mass of the installed absorbers compared with the host rail system, together with the fact that the objective of this section is to compare the controlled and uncontrolled rail response using the experimentally validated low-order rail model rather than to reconstruct the full absorber-resolved dynamics. Based on the given relationships, the spectral density function for all four outputs is expressed in Figure \ref{fig:9}.
The mean square of the output results with and without the TMD is shown and compared in Table \ref{table6}. Based on this idealized broadband random-excitation model, the proposed bi-resonator TMD yields an average reduction of approximately 12\% in the RMS response of the rail. Although this result should not be interpreted as a direct prediction under actual wheel--rail operating spectra, it indicates that the absorber retains a measurable vibration-reduction effect under stochastic broadband loading in addition to the deterministic impact conditions examined experimentally.
\begin{table}[htbp]
    \centering
    % Set the space between columns. Adjust '12pt' for more or less space.
    \setlength{\tabcolsep}{12pt}

    \caption{Performance of the TMD on random vibrations of railway}
    \label{tab:dva_performance}
    \begin{tabular}{@{}cccc@{}}
        \toprule
        \multirow{2}{*}{$i$} & \multicolumn{2}{c}{$\dfrac{\mathrm{E}[y_i^2] \times 10^{11}}{S_0}$} & \multirow{2}{*}{Improvement (\%)} \\
        \cmidrule(lr){2-3}
        & Without TMD & With TMD & \\
        \midrule
        1 & 1.530 & 1.432 & 6.5 \\
        2 & 4.027 & 3.406 & 15.6 \\
        3 & 5.583 & 4.871 & 12.7 \\
        4 & 1.491 & 1.325 & 11.1 \\
        \bottomrule
    \end{tabular}
    \label{table6}
\end{table}
%%%%%%%%%%%%%%%%%%%%%%%%%%%%%%%%%%%%%%%%%%%%%%%%%%%%%%
\section{Conclusions}
This study presented a broadband design methodology for railway tuned mass dampers (TMDs) based on the concept of joining multiple locally resonant bandgaps. Rather than designing individual absorbers to suppress isolated resonances, the proposed framework combines multiple resonant elements within a single TMD to achieve broadband vibration attenuation while maintaining a compact and practical design suitable for railway applications. The methodology began with the experimental modal characterization of a UIC60 rail, followed by the development of an equivalent lumped-parameter model to identify the dominant vibration modes requiring suppression. Based on the proposed bandgap-joining strategy, a systematic procedure was established to design multi-resonator TMDs, and the fabricated resonators were experimentally shown to accurately reproduce their intended dynamic behavior. The proposed TMDs were subsequently evaluated on a 3-m railway track using a finite periodic arrangement of six absorbers. The experimental results demonstrated that the optimized single-resonator configuration reduced the vibration amplitudes at the dominant resonances by up to 74\%, while incorporating a second resonator further broadened the effective attenuation bandwidth up to 98\%, confirming the effectiveness of the proposed multi-resonator concept for broadband vibration suppression.

Finally, the experimentally validated rail model was employed to examine the effect of the proposed TMDs under an idealized stochastic broadband excitation. Using a band-limited white-noise input as a simplified random-loading benchmark, the analysis predicted an average reduction of approximately 12\% in the RMS vibration response. Although this result does not constitute a full vehicle--track operational prediction, it indicates that the proposed absorbers remain effective under broadband random loading in addition to deterministic impact excitation. Overall, the proposed methodology provides a practical framework for translating locally resonant metamaterial concepts into manufacturable railway vibration absorbers. Although only one- and two-resonator configurations were experimentally investigated in this work for manufacturing simplicity, the design framework is readily extendable to a larger number of internal resonators, offering a promising pathway toward compact broadband vibration mitigation systems for railway infrastructure.
\\
\bibliographystyle{IEEEtran}
\bibliography{Ref}

\appendix
\setcounter{section}{0}
\renewcommand{\thesection}{\Roman{section}}

%%%%%%%%%%%%%%%%%%%%%%%%%%%%%%%%%%%%%%%%%%%%%%%%%%%%%%
\section*{Appendix A:}

The mass and stiffness matrices of the system can be displayed as follows:
\label{Appendix A}
\begin{equation}
    \mathbf{K} =
    \begin{bmatrix}
        1154927 & -349978 & 0 & 0 \\
        -349978 & 699956 & -349978 & 0 \\
        0 & -349978 & 699956 & -349978 \\
        0 & 0 & -349978 & 1154927
    \end{bmatrix}
    \, \frac{\text{KN}}{\text{m}}
    \label{Appendix A1}
\end{equation}
\begin{equation}
    \mathbf{M} =
    \begin{bmatrix}
        41.5 & 0 & 0 & 0 \\
        0 & 41.5 & 0 & 0 \\
        0 & 0 & 41.5 & 0 \\
        0 & 0 & 0 & 41.5
    \end{bmatrix}
    \, \text{kg}
\end{equation}
The results are improved in the next step according to the stiffness modification technique. So, the following relationship needs to be considered:
\begin{equation}
    \left( [H(\omega^*)] [\kappa] + \frac{1}{\gamma_k}[I] \right) [Y] = \{0\}
\end{equation}
Procedure will be done two times with the following $\kappa$ matrices:
\begin{equation}
    \boldsymbol{\kappa} =
    \begin{bmatrix}
        0 & 0 & 0 & 0 \\
        0 & +1 & -1 & 0 \\
        0 & -1 & +1 & 0 \\
        0 & 0 & 0 & 0
    \end{bmatrix}
\end{equation}
\begin{equation}
    \boldsymbol{\kappa} =
    \begin{bmatrix}
        0 & 0 & 0 & 0 \\
        0 & 0 & 0 & 0 \\
        0 & 0 & +1 & -1 \\
        0 & 0 & -1 & +1
    \end{bmatrix}
\end{equation}
In the next step, the results are improved according to the mass modification technique. So that the following relationship needs to be considered:
\begin{equation}
    \left( \omega^2 [H(\omega^*)] [\epsilon] - \frac{1}{\zeta_m}[I] \right) [Y] = \{0\}
\end{equation}
This operation is performed three times using the following matrices:
\begin{equation}
    \boldsymbol{\epsilon} =
    \begin{bmatrix}
        0 & 0 & 0 & 0 \\
        0 & 1 & 0 & 0 \\
        0 & 0 & 1 & 0 \\
        0 & 0 & 0 & 0
    \end{bmatrix}
\end{equation}
\begin{equation}
    \boldsymbol{\epsilon} =
    \begin{bmatrix}
        1 & 0 & 0 & 0 \\
        0 & 0 & 0 & 0 \\
        0 & 0 & 0 & 0 \\
        0 & 0 & 0 & 1
    \end{bmatrix}
\end{equation}
\begin{equation}
    \boldsymbol{\epsilon} =
    \begin{bmatrix}
        0 & 0 & 0 & 0 \\
        0 & 0 & 0 & 0 \\
        0 & 0 & 1 & 0 \\
        0 & 0 & 0 & 0
    \end{bmatrix}
\end{equation}
In this case, the frequency changes of the system are as follows:
\begin{align*}
    \omega_1 &= 2858 \, \frac{\text{rad}}{\text{s}} = 455\,\text{Hz} \rightarrow \text{Error} = 1\% \\
    \omega_2 &= 3807 \, \frac{\text{rad}}{\text{s}} = 605\,\text{Hz} \rightarrow \text{Error} = 0\% \\[1.5em] % Adds extra vertical space
    \omega_3 &= 4838 \, \frac{\text{rad}}{\text{s}} = 771\,\text{Hz} \rightarrow \text{Error} = 0\% \\
    \omega_4 &= 6987 \, \frac{\text{rad}}{\text{s}} = 1112\,\text{Hz} \rightarrow \text{Error} = 13.5\%
\end{align*}
Due to a lot of trial and error, the mass modification could not be effective, and the fourth natural frequency could be modified by $\kappa$ matrix. So:

\begin{equation}
    \boldsymbol{\kappa} =
    \begin{bmatrix}
        0 & 0 & 0 & 0 \\
        0 & +1 & -1 & 0 \\
        0 & -1 & +1 & 0 \\
        0 & 0 & 0 & 0
    \end{bmatrix}
\end{equation}
The following values show updated natural frequencies:
\begin{align*}
    \omega_1 &= 2859 \, \frac{\text{rad}}{\text{s}} = 455\text{Hz}    \rightarrow \text{Error} = 1\% \\
    \omega_2 &= 3682 \, \frac{\text{rad}}{\text{s}} = 587\text{Hz} \rightarrow \text{Error} = 2.8\% \\
    \omega_3 &= 4674 \, \frac{\text{rad}}{\text{s}} = 744\text{Hz} \rightarrow \text{Error} = 3.5\% \\
    \omega_4 &= 6157 \, \frac{\text{rad}}{\text{s}} = 980\text{Hz} \rightarrow \text{Error} = 0\%
\end{align*}
The third natural frequency error will be reduced as the final step. As a result, the matrix is assumed to be as follows:
\begin{equation}
    \boldsymbol{\kappa} =
    \begin{bmatrix}
        0 & 0 & 0 & 0 \\
        0 & 0 & 0 & 0 \\
        0 & 0 & +1 & -1 \\
        0 & 0 & -1 & +1
    \end{bmatrix}
\end{equation}
The updated natural frequencies will be as follows:
\begin{align*}
    \omega_1 &= 2896 \, \frac{\text{rad}}{\text{s}} = 461\text{Hz} \rightarrow \text{Error} = 0.2\% \\
    \omega_2 &= 3688 \, \frac{\text{rad}}{\text{s}} = 587\text{Hz} \rightarrow \text{Error} = 2.8\% \\
    \omega_3 &= 4844 \, \frac{\text{rad}}{\text{s}} = 771\text{Hz} \rightarrow \text{Error} = 0\% \\
    \omega_4 &= 6195 \, \frac{\text{rad}}{\text{s}} = 986\text{Hz} \rightarrow \text{Error} = 0.6\%
\end{align*}
Finally, every frequency fell within the desired range and 3\% of the error. Therefore, the relevant results can be used to continue the calculations. Hence, the development of modifying the stiffness and mass matrices will be as follows:
\begin{equation}
    \mathbf{K} =
    \begin{bmatrix}
        1154927 & -349978 & 0 & 0 \\
        -349978 & 482998 & -133020 & 0 \\
        0 & -133020 & 293657 & -160636 \\
        0 & 0 & -160636 & 965585
    \end{bmatrix}
    \times 10^3 \, \frac{\text{N}}{\text{m}}
\end{equation}
\begin{equation}
    \mathbf{M} =
    \begin{bmatrix}
        62.6 & 0 & 0 & 0 \\
        0 & 17.57 & 0 & 0 \\
        0 & 0 & 13.06 & 0 \\
        0 & 0 & 0 & 62.6
    \end{bmatrix}
    \, \text{kg}
\end{equation}
The damping matrix is extracted to ensure sufficient stiffness and mass matrices. Therefore, Riley damping is used, and according to this method:
\begin{equation}
    C = a_0 M + a_1 K
\end{equation}
Where the values of the coefficients $a_0$ and $a_1$ are obtained by the following relation:
\begin{equation}
    \begin{Bmatrix} a_0 \\ a_1 \end{Bmatrix}
    = \frac{2\xi}{\omega_n + \omega_m}
    \begin{Bmatrix} \omega_n \omega_m \\ 1 \end{Bmatrix}
\end{equation}
Considering the median of the natural frequencies and the contribution to the attenuation that has been extracted for the vibration test, it is equal to 0.0088 can write:
\begin{equation}
    \mathbf{C} = 34.4\mathbf{M} + 2.04 \times 10^{-6} \mathbf{K} =
    \begin{bmatrix}
        4493 & -714 & 0 & 0 \\
        -714 & 2295 & -477 & 0 \\
        0 & -477 & 1192 & -276 \\
        0 & 0 & -276 & 4055
    \end{bmatrix}
    \, \frac{\text{N.s}}{\text{m}}
\end{equation}
Therefore, the modal matrix of the system can be represented as follows:
\begin{equation}
    \mathbf{\Phi} =
    \begin{bmatrix}
        0.0717 & -0.052 & 0.0788 & 0.0857 \\
        -0.187 & -0.128 & 0.0046 & 0.0762 \\
        0.0947 & -0.148 & -0.115 & -0.064 \\
        -0.033 & -0.032 & 0.071  & -0.06
    \end{bmatrix}
\end{equation}

%%%%%%%%%%%%%%%%%%%%%%%%%%%%%%%%%%%%%%%%%%%%%%%%%%%%%%
\section*{Appendix B:}

\label{Appendix B}

The derived equations \ref{eq14} and \ref{eq15} is obtained as follows:

\begin{equation}
    \begin{bmatrix} y_1(t) \\ y_2(t) \\ y_3(t) \\ y_4(t) \end{bmatrix}
    =
    \begin{bmatrix}
        \phi_{11} & \phi_{12} & \phi_{13} & \phi_{14} \\
        \phi_{21} & \phi_{22} & \phi_{23} & \phi_{24} \\
        \phi_{31} & \phi_{32} & \phi_{33} & \phi_{34} \\
        \phi_{41} & \phi_{42} & \phi_{43} & \phi_{44}
    \end{bmatrix}
    \begin{bmatrix} Z_1(t) \\ Z_2(t) \\ Z_3(t) \\ Z_4(t) \end{bmatrix}
    \label{Appendix B1}
\end{equation}
By applying the E[ ] operator on both sides of the above equation, one gets:
\begin{equation}
    \mathrm{E} \begin{bmatrix} y_1(t) \\ y_2(t) \\ y_3(t) \\ y_4(t) \end{bmatrix}
    =
    \begin{bmatrix}
        \phi_{11} & \phi_{12} & \phi_{13} & \phi_{14} \\
        \phi_{21} & \phi_{22} & \phi_{23} & \phi_{24} \\
        \phi_{31} & \phi_{32} & \phi_{33} & \phi_{34} \\
        \phi_{41} & \phi_{42} & \phi_{43} & \phi_{44}
    \end{bmatrix}
    \mathrm{E} \begin{bmatrix} Z_1(t) \\ Z_2(t) \\ Z_3(t) \\ Z_4(t) \end{bmatrix}
\end{equation}
To find the auto-correlation of outputs, we proceed as follows:
\begin{align}
    \mathrm{R}_{y_1}(\tau) &= \mathrm{E}[y_1(t) y_1(t+\tau)] \\
     \mathrm{R}_{y_2}(\tau) &= \mathrm{E}[y_2(t) y_2(t+\tau)] \\
     \mathrm{R}_{y_3}(\tau) &= \mathrm{E}[y_3(t) y_3(t+\tau)] \\
     \mathrm{R}_{y_4}(\tau) &= \mathrm{E}[y_4(t) y_4(t+\tau)]
\end{align}
\begin{equation}
    \begin{split}
    \mathrm{R}_{y_1}(\tau) = \mathrm{E}\big[ &(\phi_{11}Z_1(t) + \phi_{12}Z_2(t) + \phi_{13}Z_3(t) + \phi_{14}Z_4(t)) \\
    &(\phi_{11}Z_1(t+\tau) + \phi_{12}Z_2(t+\tau) + \phi_{13}Z_3(t+\tau) + \phi_{14}Z_4(t+\tau)) \big]
    \end{split}
\end{equation}
\begin{equation}
    \begin{split}
    \mathrm{R}_{y_2}(\tau) = \mathrm{E}\big[ &(\phi_{21}Z_1(t) + \phi_{22}Z_2(t) + \phi_{23}Z_3(t) + \phi_{24}Z_4(t)) \cdot \\
    &(\phi_{21}Z_1(t+\tau) + \phi_{22}Z_2(t+\tau) + \phi_{23}Z_3(t+\tau) + \phi_{24}Z_4(t+\tau)) \big]
    \end{split}
\end{equation}
\begin{equation}
    \begin{split}
    \mathrm{R}_{y_3}(\tau) = \mathrm{E}\big[ &(\phi_{31}Z_1(t) + \phi_{32}Z_2(t) + \phi_{33}Z_3(t) + \phi_{34}Z_4(t)) \cdot \\
    &(\phi_{31}Z_1(t+\tau) + \phi_{32}Z_2(t+\tau) + \phi_{33}Z_3(t+\tau) + \phi_{34}Z_4(t+\tau)) \big]
    \end{split}
\end{equation}

\begin{equation}
    \begin{split}
    \mathrm{R}_{y_4}(\tau) = \mathrm{E}\big[ &(\phi_{41}Z_1(t) + \phi_{42}Z_2(t) + \phi_{43}Z_3(t) + \phi_{44}Z_4(t)) \\
    &(\phi_{41}Z_1(t+\tau) + \phi_{42}Z_2(t+\tau) + \phi_{43}Z_3(t+\tau) + \phi_{44}Z_4(t+\tau)) \big]
    \end{split}
\end{equation}

\begin{equation}
    \begin{split}
    \mathrm{R}_{y_1}(\tau) = \; &\phi_{11}^2 \mathrm{R}_{Z_{11}}(\tau) + \phi_{11}\phi_{12}\mathrm{R}_{Z_{12}}(\tau) + \phi_{11}\phi_{13}\mathrm{R}_{Z_{13}}(\tau) + \phi_{11}\phi_{14}\mathrm{R}_{Z_{14}}(\tau) + \phi_{12}\phi_{11}\mathrm{R}_{Z_{21}}(\tau) \\
    + &\phi_{12}^2 \mathrm{R}_{Z_{22}}(\tau) + \phi_{12}\phi_{13}\mathrm{R}_{Z_{23}}(\tau) + \phi_{12}\phi_{14}\mathrm{R}_{Z_{24}}(\tau) + \phi_{13}\phi_{11}\mathrm{R}_{Z_{31}}(\tau) \\
    + &\phi_{13}\phi_{12}\mathrm{R}_{Z_{32}}(\tau) + \phi_{13}^2 \mathrm{R}_{Z_{33}}(\tau) + \phi_{13}\phi_{14}\mathrm{R}_{Z_{34}}(\tau) + \phi_{14}\phi_{11}\mathrm{R}_{Z_{41}}(\tau) \\
    + &\phi_{14}\phi_{12}\mathrm{R}_{Z_{42}}(\tau) + \phi_{14}\phi_{13}\mathrm{R}_{Z_{43}}(\tau) + \phi_{14}^2 \mathrm{R}_{Z_{44}}(\tau)
    \end{split}
\end{equation}

\begin{equation}
    \begin{split}
    \mathrm{R}_{y_2}(\tau) = \; &\phi_{21}^2 \mathrm{R}_{Z_{11}}(\tau) + \phi_{21}\phi_{22}\mathrm{R}_{Z_{12}}(\tau) + \phi_{21}\phi_{23}\mathrm{R}_{Z_{13}}(\tau) + \phi_{21}\phi_{24}\mathrm{R}_{Z_{14}}(\tau) + \phi_{22}\phi_{21}\mathrm{R}_{Z_{21}}(\tau) \\
    + &\phi_{22}^2 \mathrm{R}_{Z_{22}}(\tau) + \phi_{22}\phi_{23}\mathrm{R}_{Z_{23}}(\tau) + \phi_{22}\phi_{24}\mathrm{R}_{Z_{24}}(\tau) + \phi_{23}\phi_{21}\mathrm{R}_{Z_{31}}(\tau) \\
    + &\phi_{23}\phi_{22}\mathrm{R}_{Z_{32}}(\tau) + \phi_{23}^2 \mathrm{R}_{Z_{33}}(\tau) + \phi_{23}\phi_{24}\mathrm{R}_{Z_{34}}(\tau) + \phi_{24}\phi_{21}\mathrm{R}_{Z_{41}}(\tau) \\
    + &\phi_{24}\phi_{22}\mathrm{R}_{Z_{42}}(\tau) + \phi_{24}\phi_{23}\mathrm{R}_{Z_{43}}(\tau) + \phi_{24}^2 \mathrm{R}_{Z_{44}}(\tau)
    \end{split}
\end{equation}

\begin{equation}
    \begin{split}
    \mathrm{R}_{y_3}(\tau) = \; &\phi_{31}^2 \mathrm{R}_{Z_{11}}(\tau) + \phi_{31}\phi_{32}\mathrm{R}_{Z_{12}}(\tau) + \phi_{31}\phi_{33}\mathrm{R}_{Z_{13}}(\tau) + \phi_{31}\phi_{34}\mathrm{R}_{Z_{14}}(\tau) + \phi_{32}\phi_{31}\mathrm{R}_{Z_{21}}(\tau) \\
    + &\phi_{32}^2 \mathrm{R}_{Z_{22}}(\tau) + \phi_{32}\phi_{33}\mathrm{R}_{Z_{23}}(\tau) + \phi_{32}\phi_{34}\mathrm{R}_{Z_{24}}(\tau) + \phi_{33}\phi_{31}\mathrm{R}_{Z_{31}}(\tau) \\
    + &\phi_{33}\phi_{32}\mathrm{R}_{Z_{32}}(\tau) + \phi_{33}^2 \mathrm{R}_{Z_{33}}(\tau) + \phi_{33}\phi_{34}\mathrm{R}_{Z_{34}}(\tau) + \phi_{34}\phi_{31}\mathrm{R}_{Z_{41}}(\tau) \\
    + &\phi_{34}\phi_{32}\mathrm{R}_{Z_{42}}(\tau) + \phi_{34}\phi_{33}\mathrm{R}_{Z_{43}}(\tau) + \phi_{34}^2 \mathrm{R}_{Z_{44}}(\tau)
    \end{split}
\end{equation}

\begin{equation}
    \begin{split}
    \mathrm{R}_{y_4}(\tau) = \; &\phi_{41}^2 \mathrm{R}_{Z_{11}}(\tau) + \phi_{41}\phi_{42}\mathrm{R}_{Z_{12}}(\tau) + \phi_{41}\phi_{43}\mathrm{R}_{Z_{13}}(\tau) + \phi_{41}\phi_{44}\mathrm{R}_{Z_{14}}(\tau) + \phi_{42}\phi_{41}\mathrm{R}_{Z_{21}}(\tau) \\
    + &\phi_{42}^2 \mathrm{R}_{Z_{22}}(\tau) + \phi_{42}\phi_{43}\mathrm{R}_{Z_{23}}(\tau) + \phi_{42}\phi_{44}\mathrm{R}_{Z_{24}}(\tau) + \phi_{43}\phi_{41}\mathrm{R}_{Z_{31}}(\tau) \\
    + &\phi_{43}\phi_{42}\mathrm{R}_{Z_{32}}(\tau) + \phi_{43}^2 \mathrm{R}_{Z_{33}}(\tau) + \phi_{43}\phi_{44}\mathrm{R}_{Z_{34}}(\tau) + \phi_{44}\phi_{41}\mathrm{R}_{Z_{41}}(\tau) \\
    + &\phi_{44}\phi_{42}\mathrm{R}_{Z_{42}}(\tau) + \phi_{44}\phi_{43}\mathrm{R}_{Z_{43}}(\tau) + \phi_{44}^2 \mathrm{R}_{Z_{44}}(\tau)
    \end{split}
\end{equation}

The above equation will be simplified as follows:
\begin{align}
    \mathrm{R}_{y_1}(\tau) &= \sum_{i=1}^{4}\sum_{j=1}^{4} \phi_{1i} \, \phi_{1j} \, \mathrm{R}_{Z_{ij}}(\tau) \\
    \mathrm{R}_{y_2}(\tau) &= \sum_{i=1}^{4}\sum_{j=1}^{4} \phi_{2i} \, \phi_{2j} \, \mathrm{R}_{Z_{ij}}(\tau) \\
    \mathrm{R}_{y_3}(\tau) &= \sum_{i=1}^{4}\sum_{j=1}^{4} \phi_{3i} \, \phi_{3j} \, \mathrm{R}_{Z_{ij}}(\tau) \\
    \mathrm{R}_{y_4}(\tau) &= \sum_{i=1}^{4}\sum_{j=1}^{4} \phi_{4i} \, \phi_{4j} \, \mathrm{R}_{Z_{ij}}(\tau)
\end{align}

Note that in the above presentation, when $i=j$ the auto-correlation displays with one element.
\begin{equation}
    \mathrm{R}_{Z_{11}}(\tau) = \mathrm{R}_{Z_1}(\tau)
\end{equation}
Since the inputs are the same in all four nodes, there will be just one kind auto-correlation and cross-correlations of inputs will be as follows:
\begin{align}
   \mathrm{R}_{x_1}(\tau) &= \mathrm{R}_{x_2}(\tau) = \mathrm{R}_{x_3}(\tau) = \mathrm{R}_{x_4}(\tau) \\
    \mathrm{R}_{x_1x_2}    &= \mathrm{E}[x_1(t) \, x_2(t + \tau)] =\mathrm{R}_{x_1}(\tau)
\end{align}
Spectral density functions are obtained using the following relations:
\begin{align}
    \mathrm{S}_{y_1} &= \frac{1}{2\pi} \int_{-\infty}^{\infty} \mathrm{R}_{y_1} \, e^{-i\omega\tau} \,\mathrm{d}\tau \\
    \mathrm{S}_{y_2} &= \frac{1}{2\pi} \int_{-\infty}^{\infty} \mathrm{R}_{y_2} \, e^{-i\omega\tau} \,\mathrm{d}\tau \\
    \mathrm{S}_{y_3} &= \frac{1}{2\pi} \int_{-\infty}^{\infty} \mathrm{R}_{y_3} \, e^{-i\omega\tau} \,\mathrm{d}\tau \\
    \mathrm{S}_{y_4} &= \frac{1}{2\pi} \int_{-\infty}^{\infty} \mathrm{R}_{y_4} \, e^{-i\omega\tau} \,\mathrm{d}\tau
\end{align}
\begin{equation}
    \mathrm{S}_{y_1} = \frac{1}{2\pi} \int_{-\infty}^{\infty} \left( \sum_{i=1}^{4}\sum_{j=1}^{4} \phi_{1i} \, \phi_{1j} \, \mathrm{R}_{Z_{ij}}(\tau) \right) e^{-i\omega\tau} \,\mathrm{d}\tau
\end{equation}
\begin{equation}
    \mathrm{S}_{y_1} = \frac{1}{2\pi} \sum_{i=1}^{4}\sum_{j=1}^{4} \phi_{1i} \, \phi_{1j} \int_{-\infty}^{\infty} \mathrm{R}_{Z_{ij}}(\tau) e^{-i\omega\tau} \,\mathrm{d}\tau
\end{equation}
\begin{equation}
    \mathrm{S}_{y_1} = \sum_{i=1}^{4}\sum_{j=1}^{4} \phi_{1i} \, \phi_{1j} \, \mathrm{S}_{Z_{ij}}(\omega)
\end{equation}

\begin{align}
    \mathrm{S}_{y_2} &= \sum_{i=1}^{4}\sum_{j=1}^{4} \phi_{2i} \, \phi_{2j} \, \mathrm{S}_{Z_{ij}}(\omega) \\
    \mathrm{S}_{y_3} &= \sum_{i=1}^{4}\sum_{j=1}^{4} \phi_{3i} \, \phi_{3j} \, \mathrm{S}_{Z_{ij}}(\omega) \\
    \mathrm{S}_{y_4} &= \sum_{i=1}^{4}\sum_{j=1}^{4} \phi_{4i} \, \phi_{4j} \, \mathrm{S}_{Z_{ij}}(\omega) 
\end{align}
Which, the $\mathrm{S}_{Z_i}$ are equal to:
\begin{equation}
    \mathrm{S}_{Z_i}(\omega) = \mathrm{S}_{x_i}(\omega) |H(\omega)|^2
\end{equation}
\begin{equation}
    \mathrm{S}_{Z_i}(\omega) = \frac{0.785 S_0}{\pi k_i^2} \times \left( \left(1 - \frac{m_i \omega^2}{k_i}\right)^2 + \left(\frac{c_i \omega}{k_i}\right)^2 \right)^{-1} ; \quad i = 1...4
\end{equation}
\begin{equation}
    \mathrm{S}_{y_4} = \sum_{i=1}^{4}\sum_{j=1}^{4} \phi_{4i} \, \phi_{4j} \, \mathrm{S}_{Z_{ij}}(\omega)
\end{equation}
\begin{equation}
    \mathrm{S}_{Z_{ij}}(\omega) = \sum_{n=1}^{4} \sum_{m=1}^{4} \mathrm{S}_{x_n x_m}(\omega) H_n^{(1)*} (\omega) H_m^{(2)}(\omega)
\end{equation}
In which $H_n^{(1)}$ is the frequency response of $Z_i(t)$ due to the $n^{\mathrm{th}}$ input and $H_m^{(2)}(\omega)$ is the frequency response of $Z_j(t)$ due to the $m^{\mathrm{th}}$ input. Finally the root mean square of signal can be given by:
\begin{equation}
    \mathrm{E}[y_i^2(t)] = \int_{-\infty}^{\infty} \mathrm{S}_{y_i}(\omega) \,\mathrm{d}\omega = \int_{-\infty}^{\infty} \sum_{i=1}^{4}\sum_{j=1}^{4} \phi_{1i} \, \phi_{1j} \, \mathrm{S}_{Z_{ij}}(\omega) \,\mathrm{d}\omega = \mathrm{S}_{x_i}(\omega)|H(\omega)|^2
\end{equation}
\begin{equation}
    \mathrm{E}[y_1^2(t)] = \sum_{i=1}^{4}\sum_{j=1}^{4} \phi_{1i} \, \phi_{1j} \int_{-\infty}^{\infty} \mathrm{S}_{Z_{ij}}(\omega) \,\mathrm{d}\omega
\end{equation}
Similarly:
\begin{align}
    \mathrm{E}[y_2^2(t)] &= \sum_{i=1}^{4}\sum_{j=1}^{4} \phi_{2i} \, \phi_{2j} \int_{-\infty}^{\infty} \mathrm{S}_{Z_{ij}}(\omega) \,\mathrm{d}\omega \\
    \mathrm{E}[y_3^2(t)] &= \sum_{i=1}^{4}\sum_{j=1}^{4} \phi_{3i} \, \phi_{3j} \int_{-\infty}^{\infty} \mathrm{S}_{Z_{ij}}(\omega) \,\mathrm{d}\omega \\
    \mathrm{E}[y_4^2(t)] &= \sum_{i=1}^{4}\sum_{j=1}^{4} \phi_{4i} \, \phi_{4j} \int_{-\infty}^{\infty} \mathrm{S}_{Z_{ij}}(\omega) \,\mathrm{d}\omega
\end{align}

%%%%%%%%%%%%%%%%%%%%%%%%%%%%%%%%%%%%%%%%%%%%%%%%%%%%%%

\end{document}